\documentclass[pdflatex,prd,twocolumn,showpacs,showkeys,superscriptaddress,nofootinbib]{revtex4-1}

\pdfoutput=1

\usepackage{bm,amsmath,amssymb}
\usepackage{graphicx}
\usepackage{color}
\usepackage{slashed}
\usepackage{tabularx}
\usepackage{bbold}
\usepackage{bbm}
\usepackage{placeins}
\usepackage[normalem]{ulem}
\usepackage[colorlinks=true]{hyperref}

\newcommand{\tr}{\mathrm{tr}}
\newcommand{\Tr}{\mathrm{Tr}}
\newcommand{\STr}{\mathrm{STr}}

\newcommand{\I}{\mathrm{i}}

\newcolumntype{L}[1]{>{\raggedright\arraybackslash}p{#1}} 
\newcolumntype{C}[1]{>{\centering\arraybackslash}p{#1}} 
\newcolumntype{R}[1]{>{\raggedleft\arraybackslash}p{#1}} 

\graphicspath{{.}{figures/}{mma/}}

\begin{document}

\title{Vector and Axial-Vector Mesons in Nuclear Matter}

\newcommand{\ECT}{European Centre for Theoretical Studies in Nuclear Physics and related Areas (ECT*) and Fondazione Bruno Kessler, Villa Tambosi, Strada delle Tabarelle 286, I-38123 Villazzano (TN), Italy}
\newcommand{\JLU}{Institut f\"ur Theoretische Physik, Justus-Liebig-Universit\"at Giessen, Heinrich-Buff-Ring 16, 35392 Giessen, Germany}
\newcommand{\TUDa}{Institut f\"ur Kernphysik (Theoriezentrum), Technische Universit\"at Darmstadt, Schlossgartenstr. 2, 64289 Darmstadt, Germany}
\newcommand{\GU}{Institut f\"ur Theoretische Physik, Goethe Universität, Max-von-Laue-Str.~1, 60438 Frankfurt am Main, Germany}
\newcommand{\Graz}{Institut f\"ur Physik, Karl-Franzens-Universit\"at Graz, NAWI Graz, Universit\"atsplatz~5, 8010 Graz, Austria}
\newcommand{\HFHF}{Helmholtz Forschungsakademie Hessen f\"ur FAIR (HFHF), GSI Helmholtzzentrum f\"ur Schwerionenforschung, Campus Gießen}

\author{Ralf-Arno Tripolt}\affiliation{\GU}\affiliation{\Graz}
\author{Christopher Jung}\affiliation{\JLU}
\author{Lorenz von Smekal}\affiliation{\JLU}\affiliation{\HFHF}
\author{Jochen Wambach}\affiliation{\TUDa}

\begin{abstract}
As a first step towards a realistic phenomenological description of vector and axial-vector mesons in nuclear matter, we calculate the spectral functions of the $\rho$ and the $a_1$ meson in a chiral baryon-meson model as a low-energy effective realization of QCD, taking into account the effects of fluctuations from scalar mesons, nucleons, and vector mesons within the Functional Renormalization Group (FRG) approach. The phase diagram of the effective hadronic theory exhibits a nuclear liquid-gas phase transition as well as a chiral phase transition at a higher baryon-chemical potential. The in-medium $\rho$ and $a_1$ spectral functions are calculated by using the previously introduced analytically-continued FRG (aFRG) method. Our results show strong modifications of the spectral functions in particular near the critical endpoints of both phase transitions which may well be of relevance for electromagnetic rates in heavy-ion collisions or neutrino emissivities in neutron-star merger events.
\end{abstract}



\maketitle

\section{Introduction}\label{sec:introduction}

The properties of matter under extreme conditions in temperature and/or density, as encountered for instance in the early universe, the core of neutron stars, and binary neutron star mergers are in the focus of ongoing theoretical as well as experimental and observational efforts. Hot and dense strong-interaction matter is created and studied in relativistic heavy-ion collisions at the world's most powerful accelerator facilities while the properties of neutron stars and their dynamics are inferred from observations of electroweak signals and more recently from gravitational radiation in merger events. In both cases, key challenges include the investigation of the in-medium modifications of hadrons and their connection to changes in the underlying symmetries, the equation of state, transport properties, and the phase structure of strong-interaction matter. Here, the basic features of Quantum Chromodynamics (QCD), dynamical chiral symmetry breaking and confinement, play a decisive role, as it is expected that at high temperatures and baryon densities chiral symmetry gets restored, and that quarks and gluons are liberated as confinement disappears. For overviews on heavy-ion measurements, astrophysical observations and theoretical studies, see e.g.~the reviews \cite{Stephanov:2006zvm,Braun-Munzinger:2008szb,Fukushima:2010bq,Andersen:2014xxa,Luo:2017faz,STAR:2017sal,Shuryak:2014zxa,Andronic:2017pug,Yin:2018ejt, Rezzolla:2018jee, Fischer:2018sdj, Fu:2019hdw} and references therein.
 
To determine the electroweak response of compressed and hot nuclear matter, a realistic theoretical description of the in-medium $\rho$ and $a_1$ spectral functions is required. Both are parity as well as chiral partners of the global chiral $SU(2)_L \times SU(2)_R$ symmetry of QCD for $N_f=2$ light quark flavors. Its realization therefore plays an important role for their in-medium properties. However, a consistent description that incorporates this (approximate) chiral symmetry and its spontaneous breaking as well as the effects from critical fluctuations, most notably from a critical endpoint of a possible phase transition inside a dense nuclear environment, is still missing. 

In this work we therefore present a new setup for the calculation of vector and axial-vector meson spectral functions in dense nuclear matter. Our approach is based on the Functional Renormalization Group (FRG) which represents a non-perturbative framework that is capable of including both quantum and thermal fluctuations \cite{Berges:2000ew,Polonyi:2001se,Pawlowski:2005xe,Schaefer:2006sr,Kopietz:2010zz,Braun:2011pp, Friman:2011zz, Gies:2006wv}. Most notably this includes order-parameter fluctuations due to the dynamics of collective excitations which are not accessible in mean-field approximations. This makes the FRG particularly well suited to study the critical behavior of the corresponding correlations. Moreover, as other functional methods, the FRG is not hampered by the fermion sign problem encountered in lattice QCD \cite{deForcrand:2010ys} at finite baryon density. While the region of finite temperature at low net-baryon density, for baryon chemical potentials $\mu_B$ less than about twice the temperature $T$ where lattice QCD results are available, is important to benchmark effective theories and approximations, including the unavoidable truncations in functional methods, the FRG can therefore then be applied with some confidence also in the dense region of the phase diagram of strong-interaction matter, with $\mu_B$ of the order of the nucleon mass at temperatures that are at least an order of magnitude lower than that. In addition, the necessary truncations can be made to preserve the global symmetry structure and its breaking patterns as here described by the underlying effective theory.

To construct an effective low-energy description for nuclear matter that is consistent with chiral symmetry and its breaking pattern, the notion of parity-partners in the bosonic sector has to be extended to massive fermions. This is accomplished in the parity-doublet model (PDM), or mirror baryon model \cite{Detar:1988kn,Hatsuda:1988mv,Jido:1998av}. The PDM describes nucleons along with their parity partners and can account for a finite nucleon mass in a chirally-invariant fashion. It is motivated by the assumption that a large fraction of mass of the nucleon in QCD is generated by the gluonic contribution to the scale anomaly \cite{Shifman:1978zn}, and not through dynamical chiral symmetry breaking. The PDM also provides a natural description for the parity-doubling structure of the low-lying baryons observed in recent lattice-QCD calculations \cite{Glozman:2012fj,Aarts:2017rrl}. 
The mean-field phase diagram \cite{Jido:2001nt,Wilms:2007uc,Zschiesche:2006zj,Dexheimer:2007tn,Gallas:2009qp,Sasaki:2010bp,Sasaki:2011ff,Gallas:2011qp,Steinheimer:2011ea,Paeng:2011hy,Giacosa:2011qd,Benic:2015pia} is known to consist of two distinct first-order phase transitions: the usual nuclear liquid-gas transition together with a second, chiral transition at higher chemical potentials. In particular the existence of this chiral transition inside the dense nuclear-matter phase was confirmed including fluctuations within the FRG in \cite{Weyrich:2015hha} to be a robust prediction of the PDM. Just like $\rho$ and $a_1 $, the nucleon $N(938)$ and its parity partner, commonly assigned to the $1/2^{-} $ $N^*(1535)$ resonance become (almost) degenerate at this transition, with a common and chirally invariant finite baryon mass $m_{0,N}$ from the scale anomaly. This model thus serves as  suitable effective theory to describe a chiral phase transition inside nuclear matter entirely in terms of hadronic degrees of freedom.

Experimentally testable predictions from this scenario could range from an enhanced production of $\eta $ mesons in heavy-ion collisions at low beam energies, when a population imbalance between $N$ and $N^*$ is created in such a transition, to an enhanced dilepton signal from critical fluctuations when the trajectory in the phase diagram of the expanding system comes close to the associated chiral critical endpoint. We will focus on the latter here, and calculate as a first step towards obtaining electromagnetic spectral function and thermal dilepton rates \cite{Tripolt:2018jre}, the in-medium spectral function of the $\rho$ vector meson and its chiral partner, the $a_1$ axial-vector meson in nuclear matter within this PDM setting.

For the calculation of the spectral functions we use the analytically-continued FRG (aFRG) framework developed in \cite{Kamikado:2013sia,Tripolt:2013jra, Tripolt:2014wra}. The aFRG method avoids the need for numerical reconstruction schemes, see for example \cite{Vidberg:1977,Jarrell:1996,Asakawa:2000tr,Qin:2013ufa, Dudal:2013yva,Fischer:2017kbq,Cyrol:2017ewj, Tripolt:2018xeo, Cyrol:2018xeq}, and it is thermodynamically consistent in that the thermodynamic grand potential and the spectral functions are calculated on the same footing. The aFRG method has been successfully applied in different situations, for example to calculate in-medium spectral functions of pions and the scalar $\sigma $ meson \cite{Tripolt:2013jra,Tripolt:2014wra, Tripolt:2016cey}, the quark spectral function \cite{Tripolt:2018qvi,Tripolt:2020irx} as well as vector- and axial-vector meson spectral functions at finite temperature and density in extended linear-sigma models with quarks \cite{Jung:2016yxl, Jung:2019nnr}, together with the corresponding electromagnetic spectral function and thermal dilepton rates \cite{Tripolt:2018jre} inside quark matter.

In this paper we use the PDM as our effective theory describing the two iso-doublets of $N$ and $N^*$ as parity partners with chiral representations in mirror assignment and chirally invariant mass term. These interact via chirally invariant Yukawa couplings with pions and the scalar $\sigma$ meson as well as $\rho $ and $a_1$ mesons as the respective chiral partners in the (pseudo-)scalar and the (axial-)vector meson channels. The formalism to describe the fluctuations of massive vector and axial-vector mesons in an effective theory based on (anti-)selfdual field strengths is taken from \cite{Jung:2019nnr}. Compared to the precursor study in \cite{Jung:2016yxl}, the inclusion of fluctuating vector and axial-vector mesons in the aFRG flows allows, in particular, to account for important additional contributions to their spectral functions such as, e.g., the three-body resonance decay $a_1\to \rho\pi \to 3\pi$.

Our work represents a first step towards a realistic description of vector and axial-vector mesons in nuclear matter. While the former are particularly relevant for the interpretation of dilepton spectra in heavy-ion collisions, both are needed to determine the neutrino emissivities in proto-neutron stars and binary neutron-star mergers. The in-medium spectral functions provide access to real-time quantities such as pole masses and decay widths but also to other observables such as pertinent transport coefficients. Moreover, since our FRG treatment is thermodynamically consistent and symmetry preserving, the in-medium modifications of the spectral functions can be stringently connected to the restoration of chiral symmetry.

The remainder of this paper is organized as follows. In Sec.~\ref{sec:FRG_PDM} the theoretical setup as well as results for the phase diagram and the Euclidean (screening) masses are presented. In Sec.~\ref{sec:spectral} we discuss results for the real-time two-point functions and the in-medium spectral functions of the $\rho$ and the $a_1$ meson. We close with a summary and outlook in Sec.~\ref{sec:summary}. Further details are deferred to an appendix.

\vspace{1.5cm}

\section{The parity-doublet model with vector and axial-vector mesons}
\label{sec:FRG_PDM}

\subsection{The Effective Average Action}
\label{sec:ansatz}

The question of how to describe baryons within effective models incorporating the principle of chiral symmetry has a long history. Here, important work was done by Walecka who introduced a hadronic model consisting of nucleons, scalar mesons and vector mesons \cite{Walecka:1974qa} and by Lee and Wick who reformulated the model of Walecka as a chirally invariant version \cite{Lee:1974ma}. The latter model is often called the chiral Walecka model and basically corresponds to the quark-meson model with nucleon instead of quark fields. The problem here is that the baryonic degrees of freedom become essentially massless in the chirally restored phase (Lee-Wick matter) as in these models their mass, in the chiral limit, is entirely generated by spontaneous chiral symmetry breaking. The occurrence of a Lee-Wick phase is circumvented by including the parity partners of the nucleons in a chirally invariant way, leading to parity-doublet models.

For an FRG treatment of the PDM we need an ansatz for the corresponding effective average action $\Gamma_k$, the central object in the FRG approach formulated by Wetterich \cite{Wetterich:1992yh}, where $k$ is the renormalization-group scale. In this work we will use the following ansatz for the effective average action of the PDM, extended by vector and axial-vector mesons, thus combining the FRG framework for the PDM presented in \cite{Weyrich:2015hha} and the strategy to include massive spin-1 (axial-)vector mesons presented in \cite{Jung:2019nnr},
\begin{widetext}
\begin{align}\label{eq:Gamma}
\Gamma_{k}=
\int d^{4}x \:\Big\{&
\bar{N_1}\left(\slashed\partial-\mu_B \gamma_0+h_{s,1}(\sigma+i\vec{\tau}\cdot\vec{\pi}\gamma^{5})
+h_{v,1}(\gamma_\mu \vec{\tau}\cdot\vec{\rho}_\mu+\gamma_\mu\gamma^5\vec{\tau}\cdot{\vec a}_{1,\mu})\right)N_1\nonumber\\
&+\bar{N_2}\left(\slashed\partial-\mu_B \gamma_0+h_{s,2}(\sigma-i\vec{\tau}\cdot\vec{\pi}\gamma^{5})
+h_{v,2}(\gamma_\mu \vec{\tau}\cdot\vec{\rho}_\mu-\gamma_\mu\gamma^5\vec{\tau}\cdot{\vec a_{1,\mu}}\right)N_2+m_{0,N}\left(\bar{N_1}\gamma^{5}N_2-\bar{N_2}\gamma^{5}N_1\right)\nonumber\\
&
+U_{k}(\phi^2)-c\sigma + \frac{1}{2} (D_\mu \phi)^\dagger D_\mu \phi - \frac{1}{4} \,\tr\, \partial_\mu \rho_{\mu\nu} \partial_\sigma \rho_{\sigma\nu} +\frac{m_v^2}{8} \, \tr\, \rho_{\mu\nu}\rho_{\mu\nu}
\Big\} \, .
\end{align}
\end{widetext}

The nucleon fields $N_1$ and $N_2$ are defined to have opposite parity and respectively represent the iso-doublet of nucleons, $(p,n)$, and their parity partners, to which we assign the $N^*(1535)$. The chirally-invariant bare nucleon mass is given by $m_{0,N}$; $\mu_B$ denotes the baryon chemical potential, and the $h$'s label the various Yukawa couplings between mesons and baryons. In this work we choose the scalar and vector-couplings to be the same, i.e.~$ h_{s,1}=h_{v,1}$ and $h_{s,2}=h_{v,2}$, as also done in \cite{Jung:2016yxl,Jung:2019nnr}. The scalar and pseudo-scalar meson fields are combined in $\phi^2=\sigma^2+\vec\pi^2$ with $\phi=(\sigma,\vec{\pi})^T$. $U_{k}(\phi^2)$ is the $O(4)$ symmetric effective potential, and the term $c\sigma$ provides the explicit chiral-symmetry breaking that arises from the small but finite current masses of the light quarks in perturbative QCD. In principle, the effective potential and hence the thermodynamic grand potential can depend on all field combinations allowed by symmetry. The guiding principle here is to include fluctuations due to collective excitations such as those of order-parameter fields as in Landau-Ginzburg-Wilson effective theories. Because neither the $\rho$ nor the $a_1$ meson are expected to develop non-vanishing expectation values in symmetric nuclear matter, their fluctuations are not included in the effective potential, at this level. 
This ansatz represents the leading order in a derivative expansion, also called local potential approximation (LPA) \cite{Litim:2001dt,Braun:2009si}. 

To describe the dynamics of massive vector and axial-vector fields and their couplings in an effective theory \cite{Jung:2019nnr}, right and left-handed vector mesons are first introduced as (anti-)selfdual field strengths $\tilde\rho^\pm_{\mu\nu} = \pm \rho_{\mu\nu}^\pm $ which transform according to the $(1,0)$ and $(0,1)$ representations of the Euclidean $O(4)$ replacing  the proper orthochronous Lorentz group for massive spin-1 particles, with $(1,0) \leftrightarrow (0,1)$ under parity,
\begin{equation}
\rho_{\mu\nu} = \rho_{\mu\nu}^+ + \rho_{\mu\nu}^- = {\vec\rho}^{\, +}_{\mu\nu} \cdot \vec T_R + {\vec\rho}^{\, -}_{\mu\nu} \cdot \vec T_L \,.
\end{equation} 
Here, $\vec T_{R} $ and $\vec T_{L} $ denote the $\mathfrak{so}(4)$ Lie algebra matrices, 
see App.~\ref{app:explicit} for explicit expressions and conventions, of the generators 
of the chiral $SU(2)_L \times SU(2)_R$ in the adjoint representation.\footnote{With 
$SU(2)_L \times SU(2)_R \sim SO(4)$ for all mesonic representations.}

The iso-triplet vector $\vec \rho_\mu $ and axial-vector ${\vec a}_{1\mu} $ fields with common mass $m_v$ are then obtained from these field strengths as
\begin{align}
     \vec \rho_\mu &= \frac{1}{2m_v} \, \tr\big(\partial_\sigma \rho_{\sigma\mu} \vec T_V\big) 
    \, , \label{eq:rhodef}\\
     \vec a_{1\mu} &= \frac{1}{2m_v} \, \tr\big(\partial_\sigma \rho_{\sigma\mu} \vec T_A\big)     \, , \label{eq:a1def}
\end{align}
where $\vec T_V = \vec T_R + \vec T_L $ and  $\vec T_A = \vec T_R - \vec T_L $. These represent conserved four-vector fields by construction,
\begin{align}
    \partial_\mu \vec\rho_\mu = \partial_\mu\vec a_{1\mu} = 0\,,
\end{align}
and, in particular, no $\pi \!-\! a_1$ mixing arises because the direct derivative coupling of the pion $\propto \partial_\mu \pi$ with the conserved $\vec a_{1\mu}$ field vanishes \cite{Ecker:1989yg}.

The interactions of the (axial-)vector fields, combined in the $\mathfrak{so}(4)$ matrix $V_\mu = \vec{\rho}_\mu \vec{T}_V +\vec{a}_{1\mu}\vec{T}_A$, with the $SO(4)$ vector of scalar and pseudo-scalar $\sigma$ and $\vec \pi$ mesons are determined from minimal coupling with $D_\mu=\partial_\mu+igV_\mu$, see App.~\ref{app:explicit}. 


\subsection{Flow of the Effective Potential and Numerical Implementation}
\label{sec:flow_pot}

The ansatz for the effective average action $\Gamma_k$ formulated in Eq.~(\ref{eq:Gamma}) is now used in the Wetterich equation \cite{Wetterich:1992yh} which defines the `flow' of $\Gamma_k$ and is given by 
\begin{align}
\label{eq:Wetterich}
\partial_k \Gamma_k=
&\frac{1}{2}\,\STr \left[\partial_k R_k \left(\Gamma_{k}^{(2)}+R_k\right)^{-1}\right],
\end{align}
where $R_k$ is a regulator function that suppresses momentum modes with momenta smaller than $k$, $\Gamma_{k}^{(2)}$ is the second functional derivative with respect to the fields, and the supertrace runs over all internal indices, over bosonic and fermionic field space, in momentum space including an integration over internal momenta or thermal Matsubara sums as well as the fermionic minus signs and factors of two. At the ultraviolet (UV) scale $k=\Lambda$, $\Gamma_k$ is essentially given by the bare action. By solving the Wetterich equation and lowering the scale $k$ the effects of quantum and thermal fluctuations are gradually included until the full effective action $\Gamma=\Gamma_{k=0}$ is obtained in the limit $k\rightarrow0$.

The regulator function $R_k$ has to be chosen appropriately for different types of fields \cite{Pawlowski:2015mlf}. In this work we use three-dimensional regulator functions that only regulate spatial momenta but not the energy components, at the expense of slightly breaking the Euclidean $O(4)$ symmetry \cite{Kamikado:2013sia}. While in principle also four-dimensional regulator functions can be used \cite{Pawlowski:2015mia,Pawlowski:2017gxj}, the three-dimensional regulators allow to analytically perform the integration over the internal energy component, or the corresponding Matsubara sum at finite temperature, as included in the supertrace of Eq.~(\ref{eq:Wetterich}). This in turn allows to apply the aFRG analytic continuation procedure as necessary for the calculation of the real-time two-point functions and spectral functions in the following. Explicit expressions for the different regulator functions are given in App.~$\ref{app:explicit}$.

\begin{table*}[t]
	\centering
	\begin{tabular}{C{1.4cm}|C{1.2cm}|C{1.6cm}|C{1.8cm}||C{1.2cm}|C{1.3cm}|C{1.3cm}||C{1.3cm}|C{1.0cm}|C{1.0cm}|C{1.0cm}|C{1.0cm}}
		$b_1$ [$\Lambda^2 $]  & $b_2$ & $b_3$ [$ \Lambda^{-2} $]& $c$ [$\Lambda^3$]  & $m_{0,N}$& $h_{s,1}$& $h_{s,2}$ & $f_\pi\equiv\sigma_0$ & $m_\pi$ & $m_\sigma$  &  $m_{N_1}$  &  $m_{N_2}$  \\
		&  & & &  [MeV]&$=h_{v,1}$ & $=h_{v,2}$& [MeV] & [MeV] & [MeV] & [MeV] & [MeV] 
		\\
		\hline\hline
		$\Big.$ 0.395189  & -4.66855 & 52.3117  & 1.74303$\cdot 10^{-3}$  & 800 & 6.94073 &13.3493& 92.8&137&474&938&1533
	\end{tabular}
	\caption{Parameters used for the effective potential at the UV cutoff $\Lambda = 1 $~GeV, the bare nucleon mass and the Yukawa couplings, as well as the resulting values for pion decay constant and Euclidean particle mass parameters in the IR.}
	\label{tab:pot_params} 
\end{table*}

When inserting the ansatz (\ref{eq:Gamma}) into the Wetterich equation (\ref{eq:Wetterich}), one obtains the flow equation for the effective potential,
\begin{widetext}
\begin{align}
\label{eq:flow_pot}
\partial_k U_k =\frac{k^4}{12\pi^2}\Big\{&
\frac{1+2n_B(E_{\sigma,k})}{E_{\sigma,k}}
+\frac{3(1+2n_B(E_{\pi,k}))}{E_{\pi,k}}
+\frac{4N_f}{E_{N_1,k}E_{N_2,k}}\Big[-(E_{N_1,k}+E_{N_2,k})
+E_{N_2,k}n_F(E_{N_1,k}-\mu_B)\nonumber\\
&+E_{N_1,k}n_F(E_{N_2,k}-\mu_B)
+E_{N_2,k}n_F(E_{N_1,k}+\mu_B)+E_{N_1,k}n_F(E_{N_2,k}+\mu_B)
\Big]
\Big\}.
\end{align}
\end{widetext}
Therein, we introduced  the number of flavors $N_f=2$, the bosonic and fermionic occupation numbers, $n_B$ and $n_F$, as given explicitly in App.~$\ref{app:explicit}$, and the scale-dependent particle energies for the sigma meson, the pion as well as for the nucleon and its party partner. The effective quasi-particle energies are defined as
\begin{align}
\label{eq:energies}
E_{\alpha,k}\equiv\sqrt{k^2+m_{\alpha,k}^2}, \qquad \alpha \in
\{\pi,\sigma,N_1, N_2\}\,, 
\end{align}
with the effective masses of the pion an the sigma meson given by
\begin{alignat}{2}
\label{eq:masses1}
&m_{\pi,k}^2&&=2U_k',\\
&m_{\sigma,k}^2&&=2U_k'+4 U_k''\phi_0^2,
\end{alignat}
where primes denote derivatives with respect to the chirally invariant $\phi^2 \equiv \sigma^2+\vec{\pi}^2$, and $\phi_0^2=\sigma_0^2$ is the global minimum of the effective potential in the IR. The masses of the nucleon and its party partner are defined by the eigenvalues of the mass matrix
\begin{align}
M=\begin{pmatrix}
h_{s,1}\sigma_0 \mathbb{1} \ & m_{0,N}\gamma_5\\
-m_{0,N}\gamma_5 & h_{s,2}\sigma_0 \mathbb{1}
\end{pmatrix},
\end{align}
and are given explicitly by
\begin{align}
\label{eq:mB}
m_{N_1}^2=&
\frac{1}{2}\Big(+(h_{s,1}-h_{s,2})\sigma_0 \\
  &\hskip 2cm +\sqrt{4m_{0,N}^2+\sigma_0^2(h_{s,1}+h_{s,2})^2}\Big), \nonumber\\
m_{N_2}^2=&
\frac{1}{2}\Big(-(h_{s,1}-h_{s,2})\sigma_0 \\
  &\hskip 2cm +\sqrt{4m_{0,N}^2+\sigma_0^2(h_{s,1}+h_{s,2})^2}\Big). \nonumber
\end{align}
As expected, for $\sigma_0\rightarrow 0$ the two masses become degenerate and reduce to the bare mass $m_{0,N}$, while for $m_{0,N}=0$ the model reduces to a sum of two independent fermion-meson models with masses for the nucleon and its parity partner given by $h_{s,1}\sigma$ and $h_{s,2}\sigma$. In this case the fermion masses are generated by spontaneous chiral symmetry breaking as in the Lee-Wick model \cite{Lee:1974ma}.

Note that the flow equation for the effective potential, Eq.~(\ref{eq:flow_pot}), does not depend on quantities related to vector mesons. This is because, for isospin-symmetric matter with an equal number of protons and neutrons, the isovector $\rho$ and $a_1$ mesons are not expected to develop non-vanishing expectation values and are therefore not included in the Ansatz for the effective potential. The isoscalar $\omega$ vector meson, in principle, contributes to the effective potential, see e.g.~\cite{Weyrich:2015hha,CamaraPereira:2020xla}, but has not been included here. The inclusion of the $\omega$ meson as a dynamical field with finite expectation value inside nuclear matter, and fluctuations contributing to effective potential and thermodynamics will be deferred to future work.

It remains to specify the UV initial conditions and the parameters used to solve the flow equation for the effective potential. At the UV scale $\Lambda$ we choose the effective potential to be of the form
\begin{align}
\label{eq:pot_UV}
U_\Lambda(\phi^{2}) =
b_1\phi^{2} +
b_2(\phi^{2})^{2}+
b_3(\phi^{2})^{3},
\end{align}
and then solve the flow equation numerically using the so-called ``grid method'', see for example \cite{Schaefer:2004en}. This method is based on a discretization of the field $\phi$ in $\sigma$-direction while the pion field is set to its expectation value $\langle \vec{\pi}\rangle ={0}$. Derivatives of the potential in field direction, as needed in the flow equation, are then obtained by a finite-difference scheme. We have checked explicitly that this numerical setup gives the same results as other approaches like finite-element or finite-volume methods, see also \cite{Grossi:2019urj,Koenigstein:2021syz}.

The numerical values for the parameters used for the effective potential, the bare nucleon mass, and the Yukawa couplings are summarized in Tab~\ref{tab:pot_params}. These are chosen such as to reproduce phenomenologically reasonable values for the pion decay constant and the particle masses in the vacuum at $k\rightarrow 0$, where the pion decay constant $f_\pi$ is identified with the value of the sigma field at the global minimum of the effective potential. The resulting values for $f_\pi$, the pion mass $m_\pi$, the sigma mass $m_\sigma$, the nucleon mass $m_{N_1}$, and the mass of its parity partner $m_{N_2}$ are also given in Tab~\ref{tab:pot_params}. For the UV cutoff we use $\Lambda=1000$~MeV and for the IR scale $k_{\text{IR}}=40$~MeV.

\begin{figure}[b!]
	\includegraphics[width=\columnwidth]{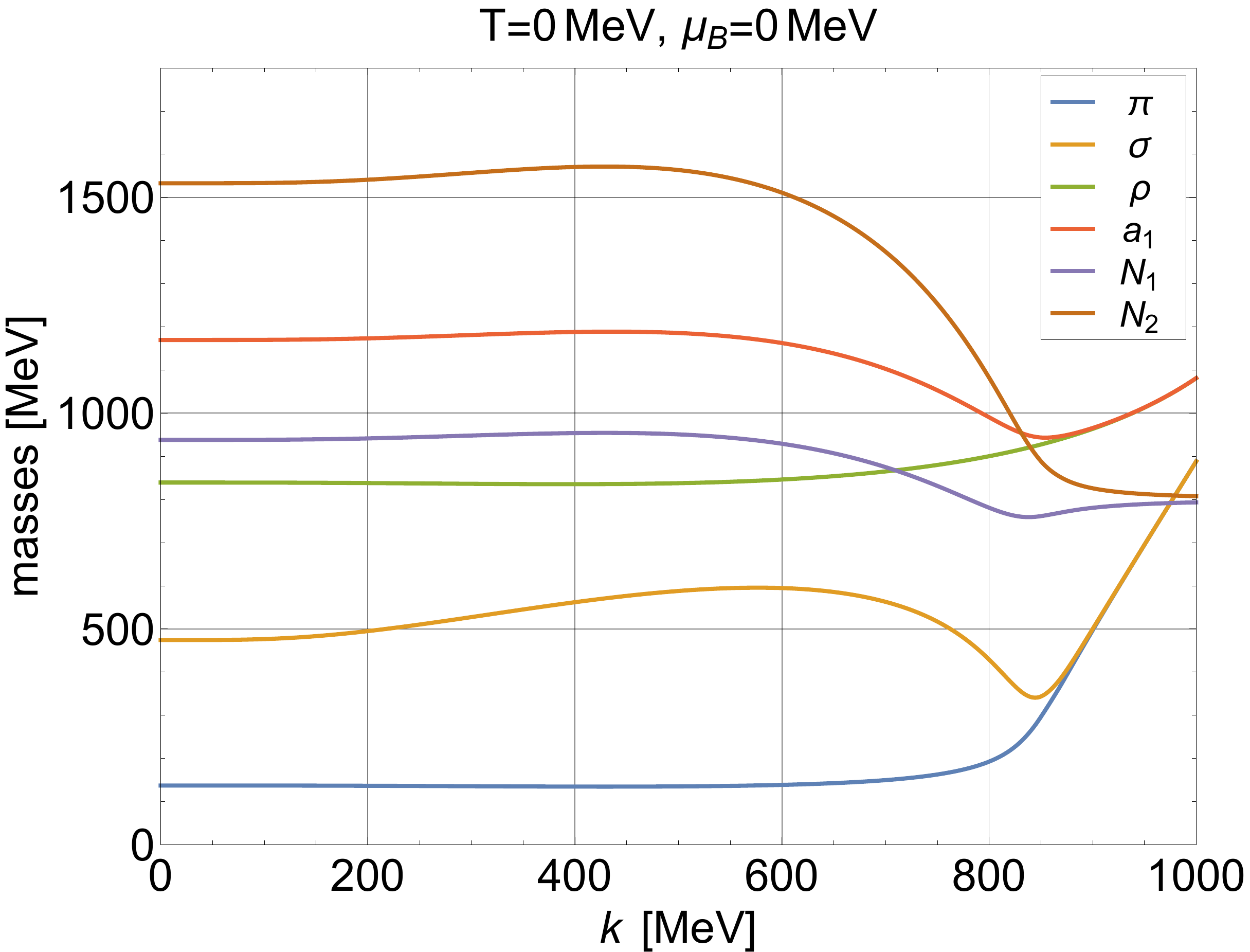}
	\caption{RG-scale dependence of the Euclidean particle masses in the vacuum. We observe strong effects from spontaneous chiral symmetry breaking between RG scales of $k\approx 900$~MeV and $k\approx 600$~MeV. The scale-dependent masses shown here serve as input for the calculation of the $\rho$ and $a_1$ spectral functions and determine the locations of the thresholds corresponding to the various decay channels.}
	\label{fig:masses_k}
\end{figure}

As usual in $O(N)$-Yukawa models, the symmetry breaking is generated from the fermionic fluctuations, where the fermionic minus sign acts like a negative index of refraction to drive the expectation value of the scalar order-parameter field away from its symmetric minimum.  Although the fermionic fluctuations arise here from baryons with a sizable bare mass of $m_{0,N} = 800 $~MeV, a UV cutoff of 
 $\Lambda = 1$~GeV turns out to be just large enough to generate the right amount of symmetry breaking starting from an effective potential with only the symmetric minimum for our choice of UV parameters. This mechanism of dynamical chiral symmetry breaking by the baryonic fluctuations is demonstrated in Fig.~\ref{fig:masses_k}. Starting from the UV cutoff $\Lambda$ the mass of the nucleons first decreases while that of the $1/2^-$ baryons increases. Incidentally, the chiral symmetry breaking scale at $k_\chi \sim 850$~MeV here coincides with the scale $m_{N_2} \sim k$ at which the heavier $1/2^-$ baryons decouple. The nucleon mass starts increasing at this point so that the fermionic fluctuations eventually cease to dominate the flow. A second scale around $k \sim 600$~MeV then emerges below which the mesonic fluctuations eventually dominate. They tend to act symmetry restoring and this thus explains why the breaking pattern is not monotonously increasing with the flow towards the infrared where it levels at the desired physical values as it would in a purely mesonic model.

The fact that this can be achieved in this way, with no clear separation of scales between the initial fermion mass of $m_{0,N} = 0.8$~GeV and the UV cutoff scale $\Lambda = 1 $~GeV might be surprising at first. It is reassuring for our effective hadronic theory which would otherwise start to lose credibility, if either considerably higher UV cutoff scales where needed or the fermionic fluctuations were irrelevant in the first place.

\subsection{Phase diagram}
\label{sec:phase_diagram}

\begin{figure}[b!]
	\includegraphics[width=\columnwidth]{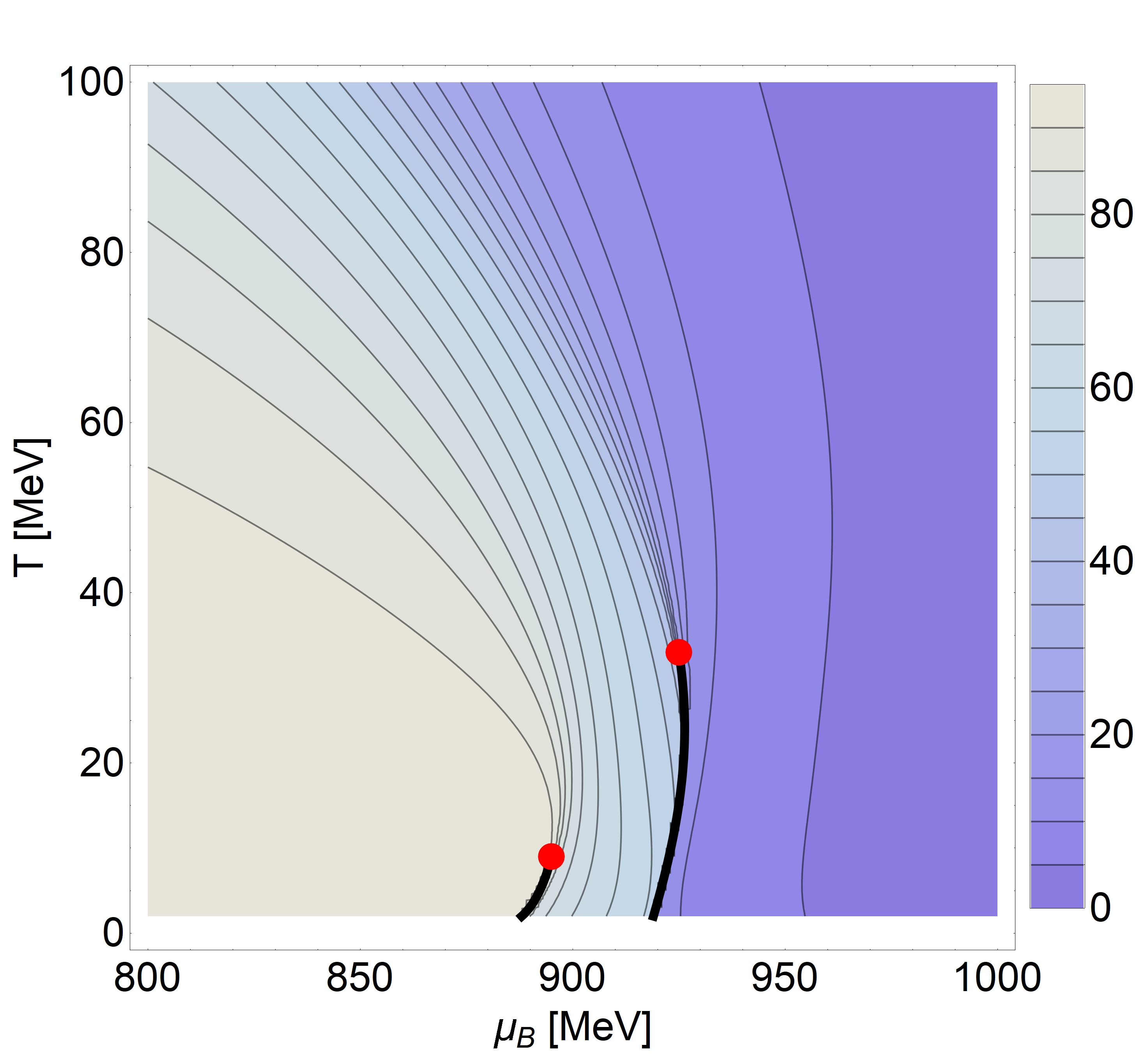}
	\caption{Phase diagram of the parity doublet model represented as a contour plot of $\sigma_0(\mu_B,T)$ with darker colors indicating smaller values, as shown in the legend bar. We observe two distinct first-order phase transitions at low temperatures which end in a critical point at ($\mu_B\approx 896$~MeV, $T\approx 10$~MeV) and at ($\mu_B\approx 925$~MeV, $T\approx 33$~MeV), respectively.}
	\label{fig:phase_diagram}
\end{figure}

In order to obtain the phase diagram of the PDM for the specified parameters we solve the flow equation for the effective potential at different combinations of temperature and baryon chemical potential, and plot the chiral order parameter, i.e.~the value of $\sigma_0(\mu_B,T)$ at the global minimum of the effective potential in the IR. The resulting phase diagram in the regime of high chemical potentials and comparatively low temperatures, as relevant for nuclear matter, is shown in Fig.~\ref{fig:phase_diagram}.

\begin{figure*}[t!]
	\includegraphics[width=0.95\textwidth]{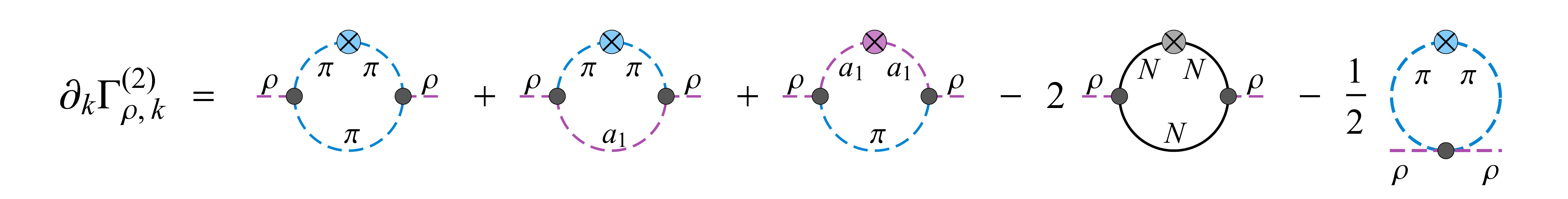}\\[-3mm]
	\includegraphics[width=0.95\textwidth]{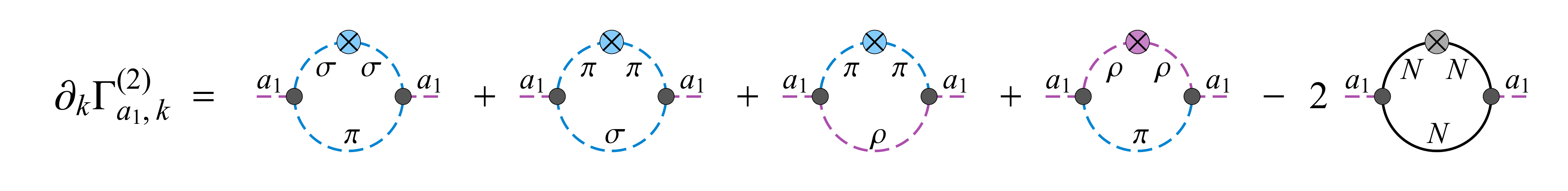}\\[-3mm]
	\includegraphics[width=0.95\textwidth]{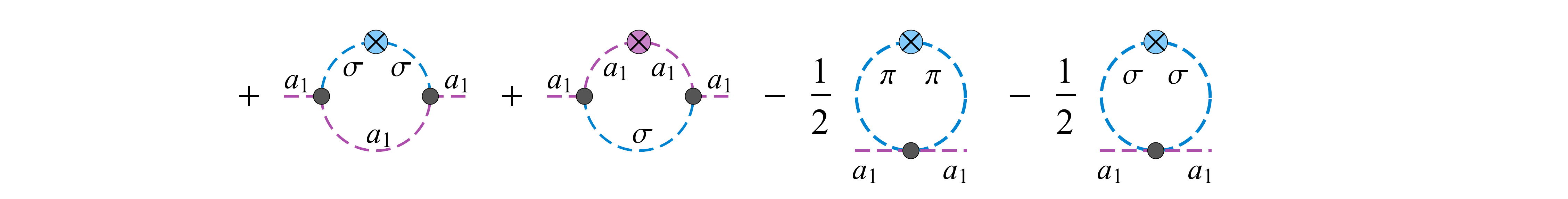}
	\caption{Flow equations of the $\rho$ and the $a_1$ two-point function in diagrammatic form. Dashed (solid) lines represent bosonic (fermionic) propagators while crossed circles indicate regulator insertions, cf.~Eq.~(\ref{eq:2PF}).}
	\label{fig:2PF}
\end{figure*}

As also found in \cite{Weyrich:2015hha}, we observe two distinct phase transitions. The phase transition at lower chemical potentials represents the liquid-gas transition of nuclear matter while the second phase transition at higher chemical potentials inside dense nuclear matter can in the chiral limit be identified as the chiral phase transition. Both phase transitions consist of a first-order line at low temperatures connected to a critical endpoint (CEP). With our current parameters these CEPs are located at ($\mu_B\approx 896$~MeV, $T\approx 10$~MeV) and ($\mu_B\approx 925$~MeV, $T\approx 33$~MeV), respectively.

The position and the strength of the liquid-gas transition strongly depend on the bare nucleon mass $m_{0,N}$. The larger the value of $m_{0,N}$ the more does the location of the discontinuity move towards larger $\mu_B$ while at the same time the strength of the transition gets weaker resulting in a larger in-medium condensate and hence a smaller nucleon sigma term. Obtaining phenomenologically acceptable values for the binding energy per nucleon, the nuclear saturation density, and the correct in-medium condensate all at the same time is known not to be possible within the present FRG setup \cite{Weyrich:2015hha}. This will require the proper inclusion of fluctuating $\omega$ mesons which we defer to a future study. For our first qualitative study here, we chose $m_{0,N}=800$~MeV as a reasonable compromise as concluded in \cite{Weyrich:2015hha}. 

At the mean-field level, the inclusion of the $\omega$ meson in the effective action 
is known to result in a simple shift of the chemical potential, and is hence effective in adjusting the binding energy per nucleon essentially without influence on the strength of the nuclear liquid-gas transition. That the same mechanism, from a mean-field gap equation for the $\omega $ meson,  does not work within the FRG framework for the order parameter fluctuations used here, was shown in \cite{Weyrich:2015hha}. The present effective theory framework for fluctuating vector mesons from \cite{Jung:2019nnr} can in principle be extended to include also the repulsive contributions from fluctuating $\omega $ mesons in the flow for the effective potential and hence the thermodynamic grand potential to improve this situation. This requires further technical developments, however, and is therefore left for future work. Another issue is the slope of the first-order lines at low temperatures. As pointed out in \cite{Tripolt:2017zgc}, from a Clausius-Clapeyron relation, a positive slope $dT_c/d\mu_c$ of the first-order line implies a negative jump in the entropy density when going from the gaseous to the liquid phase. While this by itself is not necessarily unphysical at finite temperature, when the magnitude of the jump gets too large it leads to negative entropy densities on the liquid side of the transition line which then certainly is unphysical. The inclusion of a scale-dependent gap equation for a mean-field description of the $\omega$ meson was recently shown to be able to remedy the analogous unphysical effect in the quark-meson model \cite{CamaraPereira:2020xla}. It might therefore be reasonable to expect that it will also be affected when the $\omega$-meson fluctuations are properly included within our FRG framework for a more realistic description of the thermodynamics of nuclear matter from the PDM in the future, as discussed above.

\subsection{Vector and Axial-Vector Meson Propagators and Masses}
\label{sec:vector_masses}

We now turn to the calculation of the Euclidean vector meson masses, which will use the scale-dependent effective potential as input. In Ref.~\cite{Jung:2019nnr} it was shown that the vector-meson part of the effective action in Eq.~\eqref{eq:Gamma}, 
\begin{equation}
  \label{eq:free_mvm}
  \mathcal L_0^\rho = 
 - \frac{1}{4} \,\tr\, (\partial_\mu \rho_{\mu\nu} ) \partial_\sigma \rho_{\sigma\nu} +\frac{m_v^2}{8} \, \tr\,  \rho_{\mu\nu} \rho_{\mu\nu} \,,  
\end{equation}
with Eqs.~\eqref{eq:rhodef} and \eqref{eq:a1def} corresponds to tree-level two-point functions for free (axial-)vector mesons with mass $m_v $ of the form,
\begin{equation}
  \Gamma_{\mu\nu}^{(2)}(p) = \, -\frac{m_v^2}{p^4} (p^2+m_{v}^2)\,
  \big(p^2 \delta_{\mu\nu} - p_\mu p_\nu\big) \, . \label{eq:G2Vects}
\end{equation}
It was furthermore explicitly verified that this form, upon analytic continuation, in the interacting theory correctly describes that of the corresponding single-particle contributions to the spectral representations of the propagators of massive (axial-)vector fields which are related to the analogous current-current correlation functions by current-field identities \cite{Jung:2019nnr}.

For the inclusion of (axial-)vector fluctuations of this form within the FRG we also follow the strategy of Ref.~\cite{Jung:2019nnr} and temporarily add artificial longitudinal terms in order to be able to invert the correlation functions. This then leads to an ansatz for the scale-dependent (axial-)vector propagators which reads as follows,
\begin{align}
\label{eq:vector_prop}
D_{\mu\nu,k}(p)&\equiv \left(\Gamma^{(2)}_k(p)+R_k(p)\right)_{\mu\nu}^{-1}\\
&=\frac{-p^2}{m_{0,k}^2}\frac{1}{(p^2(1+r(y))+m_{v,k}^2)}\Pi_{\mu\nu}^T(p)\nonumber\\
&\quad+\frac{-p^2}{m_{0,k}^2}\frac{1}{(p^2(1+r(y))+\xi\frac{\Lambda^2}{k^2}m_{v,k}^2)}\Pi_{\mu\nu}^L(p),
\end{align}
with a regulator shape function $r(y)$ and $y=p^2/k^2$ as defined in App.~\ref{app:explicit}, the transverse and longitudinal projection operators,
\begin{align}
\label{eq:projectors}
\Pi_{\mu\nu}^T(p)&=\delta_{\mu\nu}-p_\mu p_\nu /p^2,\\
\Pi_{\mu\nu}^L(p)&=\delta_{\mu\nu}-\Pi_{\mu\nu}^T(p)=p_\mu p_\nu /p^2,
\end{align}
and a scale-dependent mass parameter $m_{0,k}^2 \equiv  Z_k^{-1}  m_{v,k}^2 $ which 
differs from the running vector-meson pole mass $m_{v,k} $ in that it 
includes an equally scale-dependent wave-function renormalization factor $Z_k $. At the UV cutoff, we start with $Z_\Lambda = 1$ and typically $m_{0,\Lambda} = m_{v,\Lambda} \approx \Lambda $, i.e. the common pole mass of the transverse vector and axial-vector fluctuations starts out at an initial value of the same order as $\Lambda $ in the UV. The UV mass of the corresponding longitudinal fluctuations, $ \xi m_{v,\Lambda}^2$, is of the same order. The dimensionless parameter $\xi$ can be introduced to further suppress these initial longitudinal fluctuations, it is here chosen as $\xi=10 $. Because of the additional factor of $\Lambda^2/k^2$ this longitudinal mass then quickly increases with lowering $k$ and the unphysical longitudinal fluctuations decouple. Varying the parameter $\xi $, one furthermore verifies that the results are in fact completely independent of these minute longitudinal fluctuations which strictly speaking  violate the current conservation laws and require modified Ward identities at finite $k$.  
In the limit $k\rightarrow 0$ the propagators become purely transverse again, however, and thus fulfill the usual Ward identity
\begin{align}
\label{eq:ward}
p_\mu D_{k\rightarrow 0}^{\mu \nu}=0 \, .
\end{align}
For further details, see Ref.~\cite{Jung:2019nnr}.
Explicit expressions for the transverse and longitudinal regulators are given in App.~\ref{app:explicit}.

\begin{figure*}[t!]
	\includegraphics[width=\columnwidth]{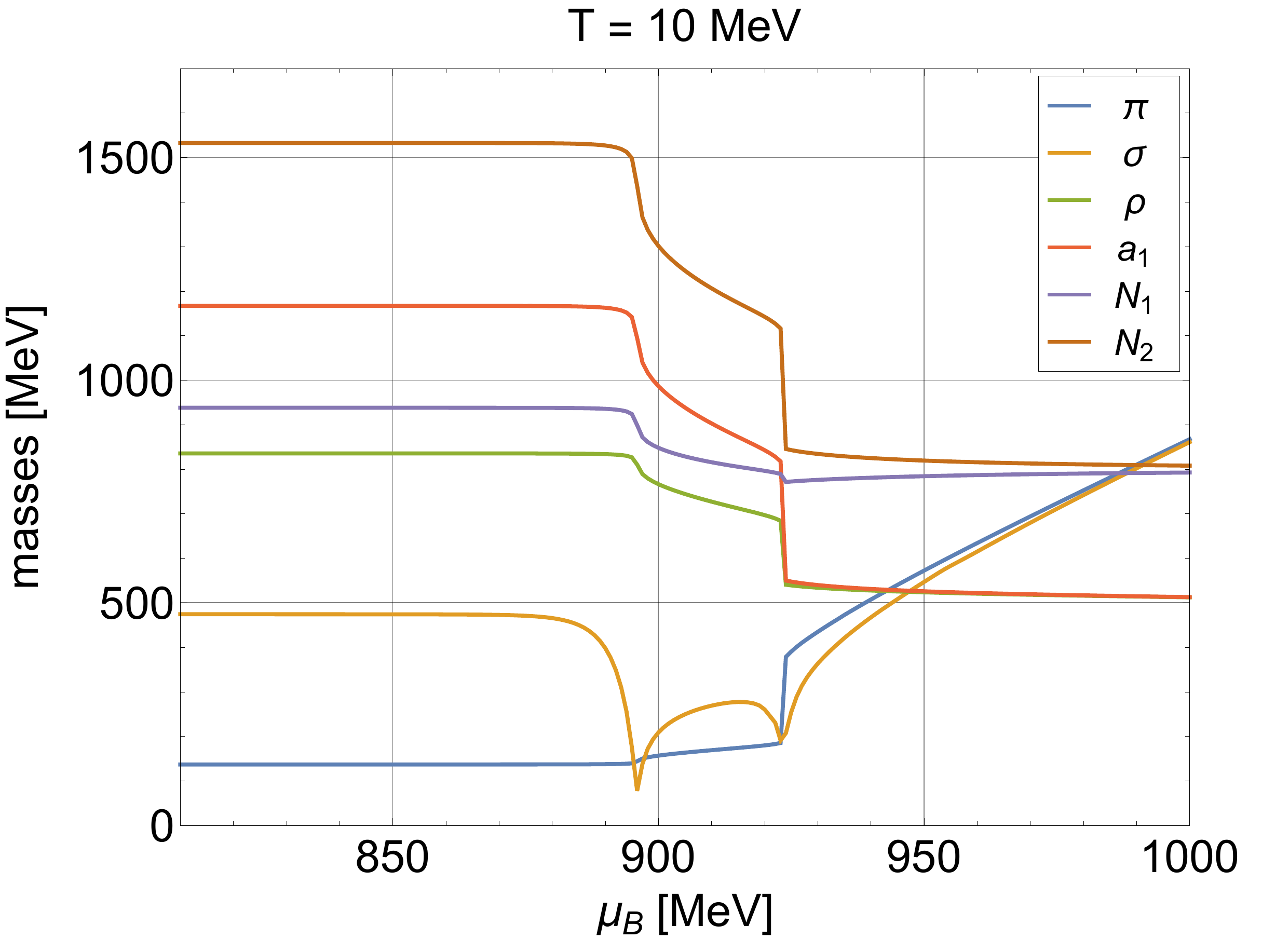}
	\includegraphics[width=\columnwidth]{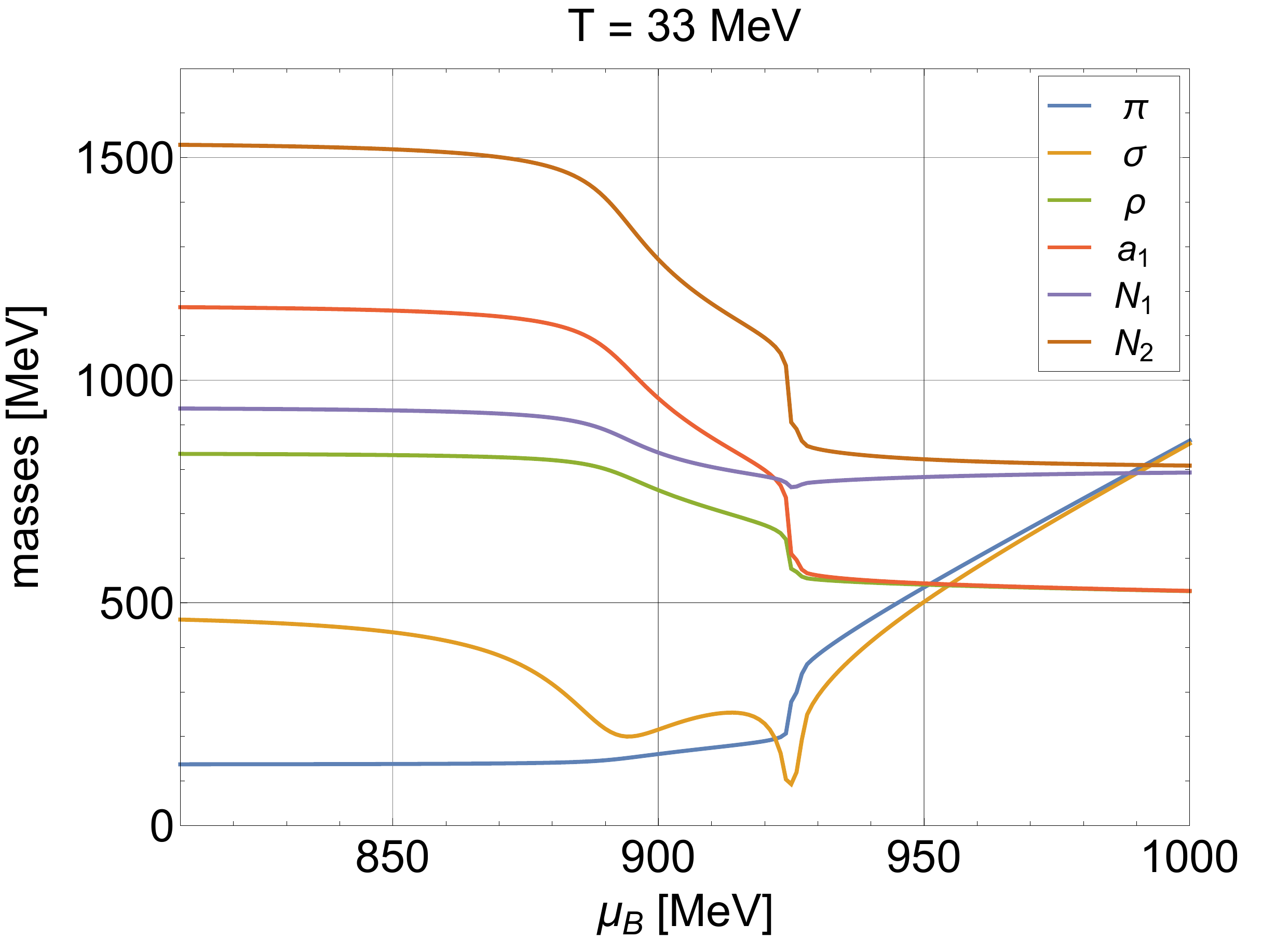}
	\caption{Euclidean particle masses as a function of baryon-chemical potential, $\mu_B$, at fixed temperatures of $T=10$~MeV (left) and $T=33$~MeV (right). Left: We observe the effects of the liquid-gas CEP at $\mu_B\approx 896$~MeV, where the sigma mass decreases rapidly, followed by the discontinuities generated by the first-order phase transition at higher chemical potentials. Right: The liquid-gas CEP still affects the behavior of the masses in a crossover from low to high density at $\mu_B\approx890$~MeV while the chiral CEP gives rise to strong modifications at $\mu_B\approx 925$~MeV. At both temperatures chiral symmetry is restored to a large extent for chemical potentials beyond the second phase transition, giving rise to degenerate masses of chiral partners.}
	\label{fig:masses_mu}
\end{figure*}

In order to calculate the masses of the (axial-)vector mesons we first need to solve the flow equation for the scale-dependent vector mass $m_{v,k}$, from which we obtain the $\rho$ and $a_1$ masses as
\begin{align}
\label{eq:masses2}
m_{\rho,k}^2&=m_{k,v}^2,\\
m_{a_1,k}^2&=m_{k,v}^2 + g^2 \phi_0^2.
\end{align}
The flow of the vector mass $m_{v,k}$ can be obtained from the flow equation for $m_{0,k}$ by observing that the flow of the product of $m_{v,k}^2$ and $m_{0,k}^2$ vanishes,
\begin{align}
\partial_k \left(m_{v,k}^2\cdot m_{0,k}^2\right)=\lim_{p\rightarrow 0} \Tr \left[p^2 \Pi_{\mu\nu}^T(p) \partial_k \left(\Gamma^{(2)}_{\rho,k}(p)\right)\right]=0.
\end{align}
We thus get 
\begin{align}
\partial_k m_{v,k}^2=-\frac{m_{v,k}^2}{m_{0,k}^2}\partial_k m_{0,k}^2,
\end{align}
with the flow of $m_{0,k}$ given by
\begin{align}
\partial_k m_{0,k}^2=-\lim_{p\rightarrow 0}\frac{\partial}{\partial |\vec{p}|^2}
 \Tr \left[p^2 \Pi_{\mu\nu}^T(p) \partial_k \left(\Gamma^{(2)}_{\rho,k}(p)\right)\right]=0.
\end{align}

The required flow equations for the two-point functions can be obtained by taking two functional derivatives of the Wetterich equation with respect to the appropriate fields which leads to the general structure
\begin{align}
\label{eq:2PF}
\partial_k\Gamma_{k}^{(2)}(p)&=\STr \Big\{ (\partial_k R_k) D_k(q) \Gamma^{(3)}_k D_k(q+p)\Gamma^{(3)}_k D_k(q) \Big\}\nonumber\\
&\quad -\frac{1}{2}\STr \Big\{(\partial_k R_k) D_k(q)\Gamma^{(4)}_k D_k(q)\Big\},
\end{align}
where explicit expression for the three- and four-point vertices are given in App.~\ref{app:explicit}. The resulting flow equation for the $\rho$ and the $a_1$ two-point function is represented diagrammatically in Fig.~\ref{fig:2PF}. Here, we take all loops into account that have up to one internal vector meson, as also done in \cite{Jung:2019nnr}. This in particular gives rise to the process $a_1\rightarrow\rho+\pi$ which is important to describe the $a_1$ spectral function. We also note that in the following we will only be dealing with the transverse two-point function which can be obtained from the corresponding flow equation,
\begin{align}
\partial_k\Gamma_{k}^{(2),T}(p)&=\frac{1}{3}\Pi_{\mu\nu}^T(p)\Tr \Big\{\partial_k \Gamma_{\mu\nu,k}^{(2)}(p)\Big\}.
\end{align}

The flow equations for $m_{0,k}$ and $m_{v,k}$ are then solved using the scale-dependent effective potential and its derivatives at the IR minimum, $\sigma_{0,\text{IR}}$, as input. As initial condition we use $m_{v,\Lambda}=m_{0,\Lambda}=1081$~MeV and $g=8.78$ for the dimensionless coupling of (axial-)vector to (pseudo-)scalar mesons (in their covariant derivative). These parameters are chosen such as to obtain phenomenological values for the $\rho$ and the $a_1$ pole mass, as discussed in the following. 

Our results for the flow of the Euclidean masses of the pion, the sigma meson, the $\rho$ and the $a_1$, as well as for the nucleon and its parity partner in the vacuum are shown in Fig.~\ref{fig:masses_k}. Here we evaluate the masses at the scale-dependent minimum of the potential and not at the fixed IR minimum in order to show the effects of chiral symmetry breaking more clearly. Starting at the UV scale where chiral symmetry is restored, the masses of the chiral partners $\pi-\sigma$, $\rho-a_1$, and $N_1-N_2$ are degenerate. Taking fluctuations into account by lowering the scale $k$, we observe the effects of chiral symmetry breaking with the masses splitting up. At the IR scale these Euclidean mass parameters then arrive at the values for (pseudo)-scalar mesons and nucleons listed in Tab.~\ref{tab:pot_params}, together with $m_\rho \approx 840$~MeV and $m_{a_1} \approx 1170$~MeV for the vector and axial-vector mesons. Note that these mass parameters do not represent the physical masses of the $\rho $ and the $a_1$ which are in turn given by the pole masses obtained from their aFRG flows and result with these parameters to be $m^p_\rho \approx 775$~MeV and $m^p_{a_1} \approx 1230$~MeV, see below.

We now turn to the dependence of the Euclidean particle masses on temperature and chemical potential. In Fig.~\ref{fig:masses_mu} we show these masses at temperatures of $T=10$~MeV and $T=33$~MeV, i.e.~at the temperatures of the two CEPs, as a function of baryon chemical potential $\mu_B$. We find that the masses are almost constant for a wide range of chemical potentials at these low temperatures, as expected from the Silver Blaze property \cite{Cohen:2003kd}. Very close to the CEP at $\mu_B\approx 896$~MeV and $T\approx 10$~MeV, however, we observe drastic changes in the masses. In particular the sigma mass strongly decreases at this second-order phase transition. In principle, the sigma meson should become exactly massless here, since it is connected to the critical long-range correlations in the density fluctuations.

At $T=10$~MeV and chemical potentials larger than the critical value of the liquid-gas CEP,  one then encounters the discontinuous behavior generated by a first-order transition. At even lower temperatures this second, chiral, phase transition is stronger than the nuclear liquid-gas transition, in that the chiral condensate and the masses change by a larger amount, cf.~also Fig.~\ref{fig:phase_diagram}. We also note that chiral symmetry becomes almost completely restored here, as evident from the fact that the masses of the chiral partners become almost degenerate. At $T\approx 33$~MeV we only see smooth changes of the masses at chemical potentials near $\mu_B\approx 896$~MeV while near the second CEP at $\mu_B\approx 925$~MeV we again observe strong but continuous changes, as expected. These masses and their scale-dependence will serve as input for the calculation of the vector-meson spectral functions discussed in the following.

\section{Vector and Axial-Vector Meson Spectral Functions in Nuclear Matter}
\label{sec:spectral}

\subsection{Analytic continuation and real-time two-point functions}

\begin{figure}[b!]
	\includegraphics[width=\columnwidth]{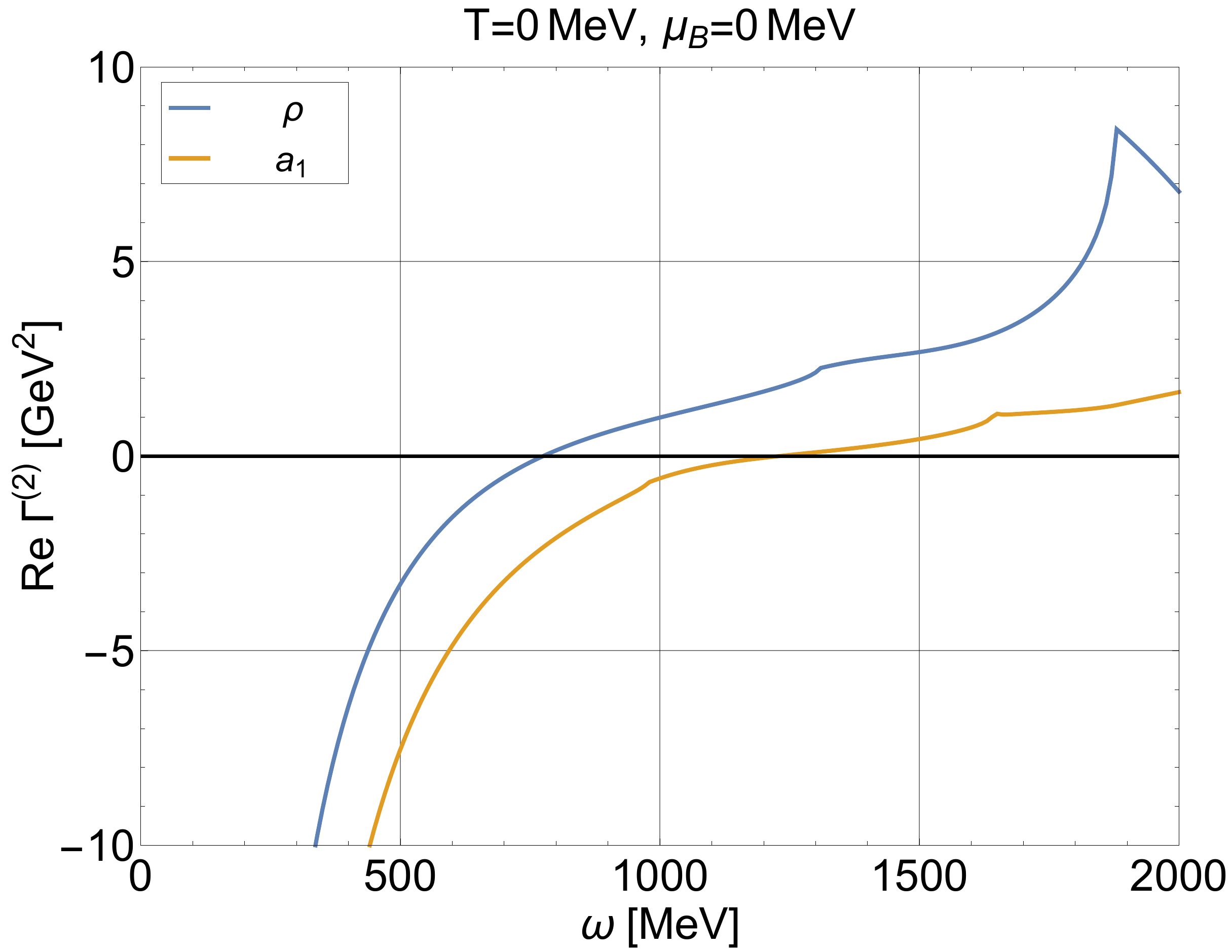}
	\caption{Real part of the $\rho$ and the $a_1$ two-point functions vs.~energy $\omega$, evaluated at $\sigma_{0,\text{IR}}$ in the vacuum, at $T=0$~MeV and $\mu_B=0$~MeV. The zero-crossings determine the respective pole masses which we find to be $m_\rho^p\approx 775$~MeV and $m_{a_1}^p\approx 1230$~MeV in the vacuum. The presence of decay channels modify the shape of the two-point functions, with the strongest effect stemming from the decay into two nucleons, $\rho^*,a_1^*\rightarrow N_1+\bar N_1$.}
	\label{fig:ReGamma2_vac}
\end{figure}

\begin{figure*}[t!]
	\includegraphics[width=\columnwidth]{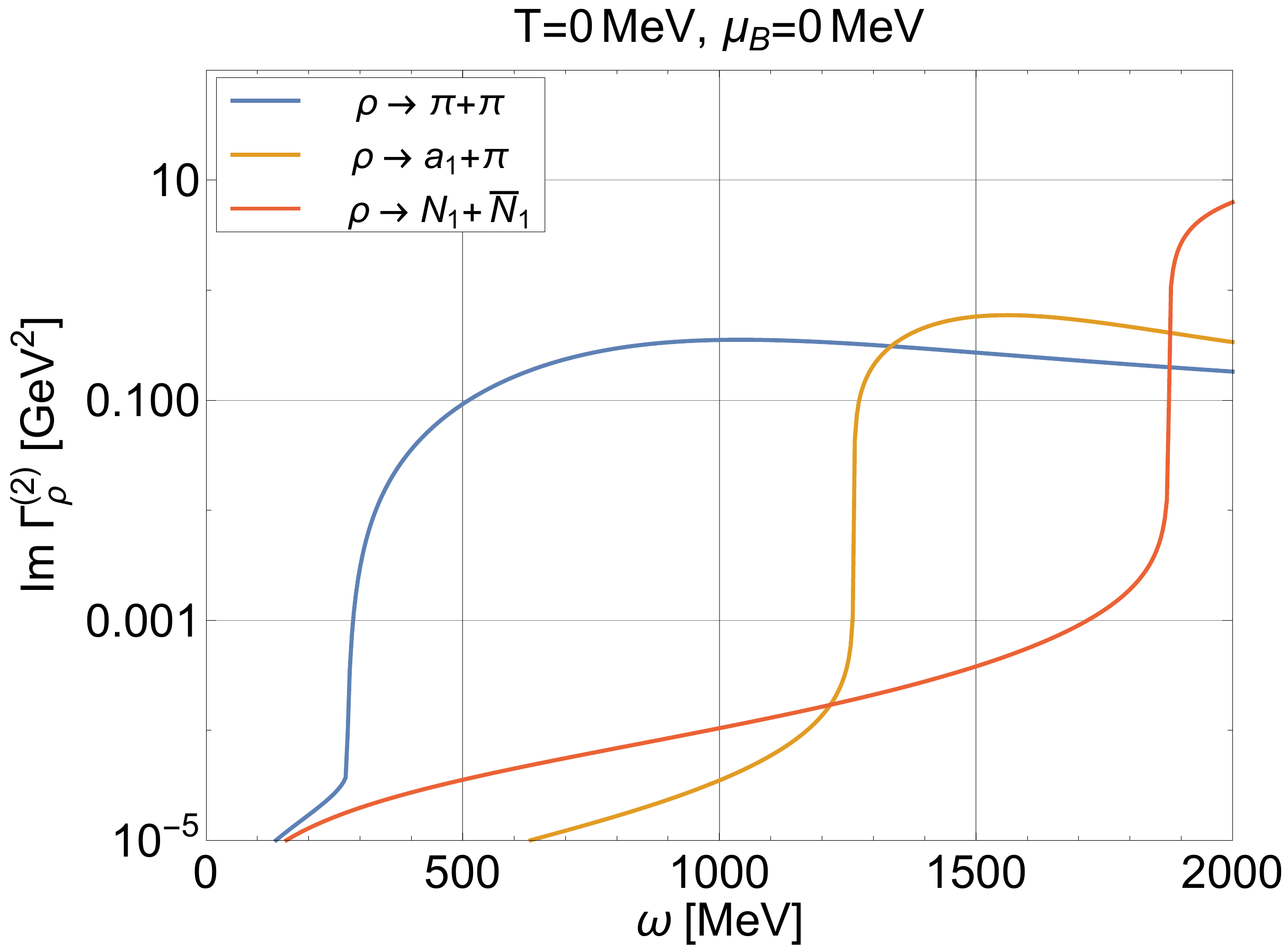}
	\includegraphics[width=\columnwidth]{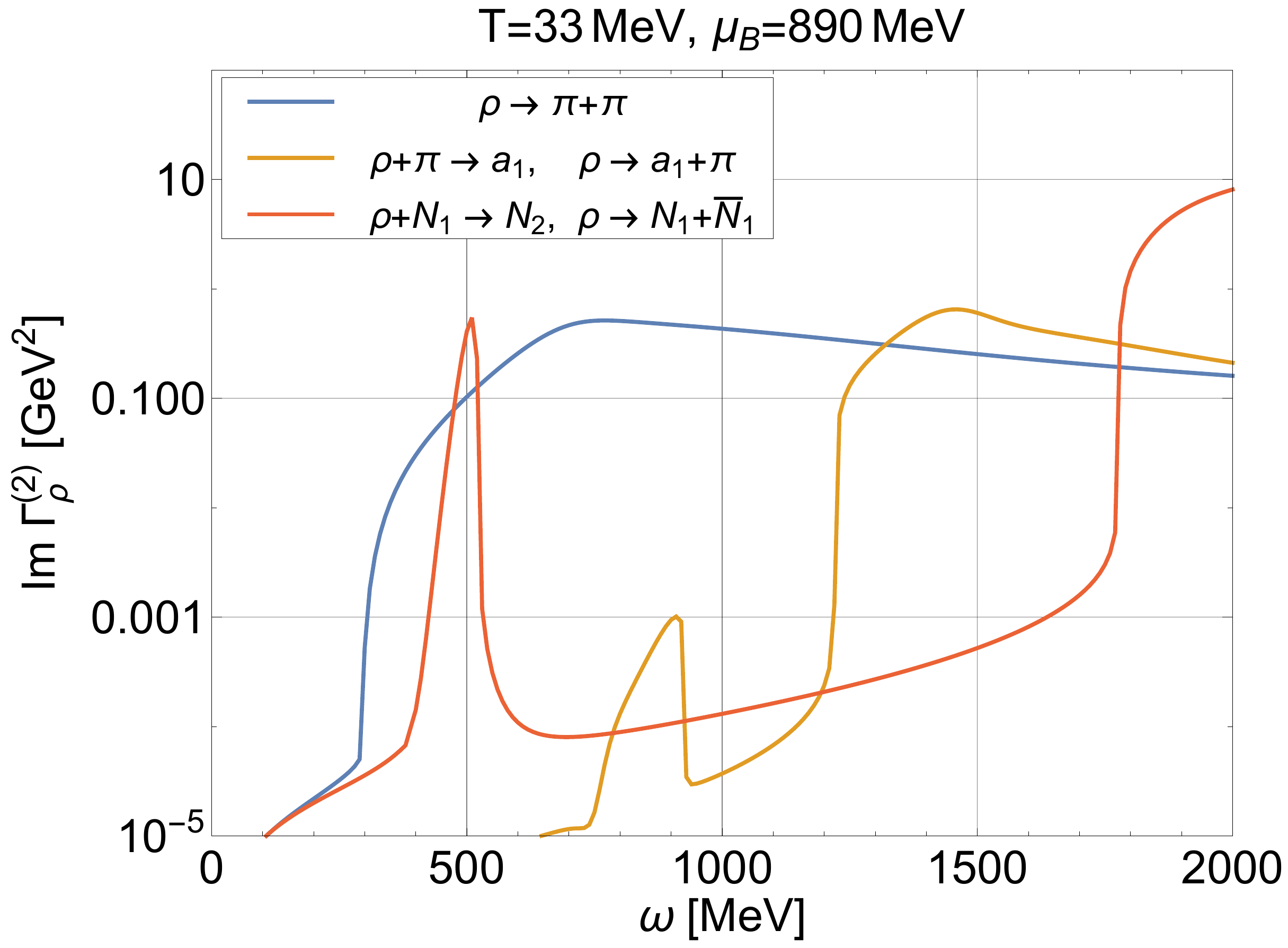}\\[2mm]
	\includegraphics[width=\columnwidth]{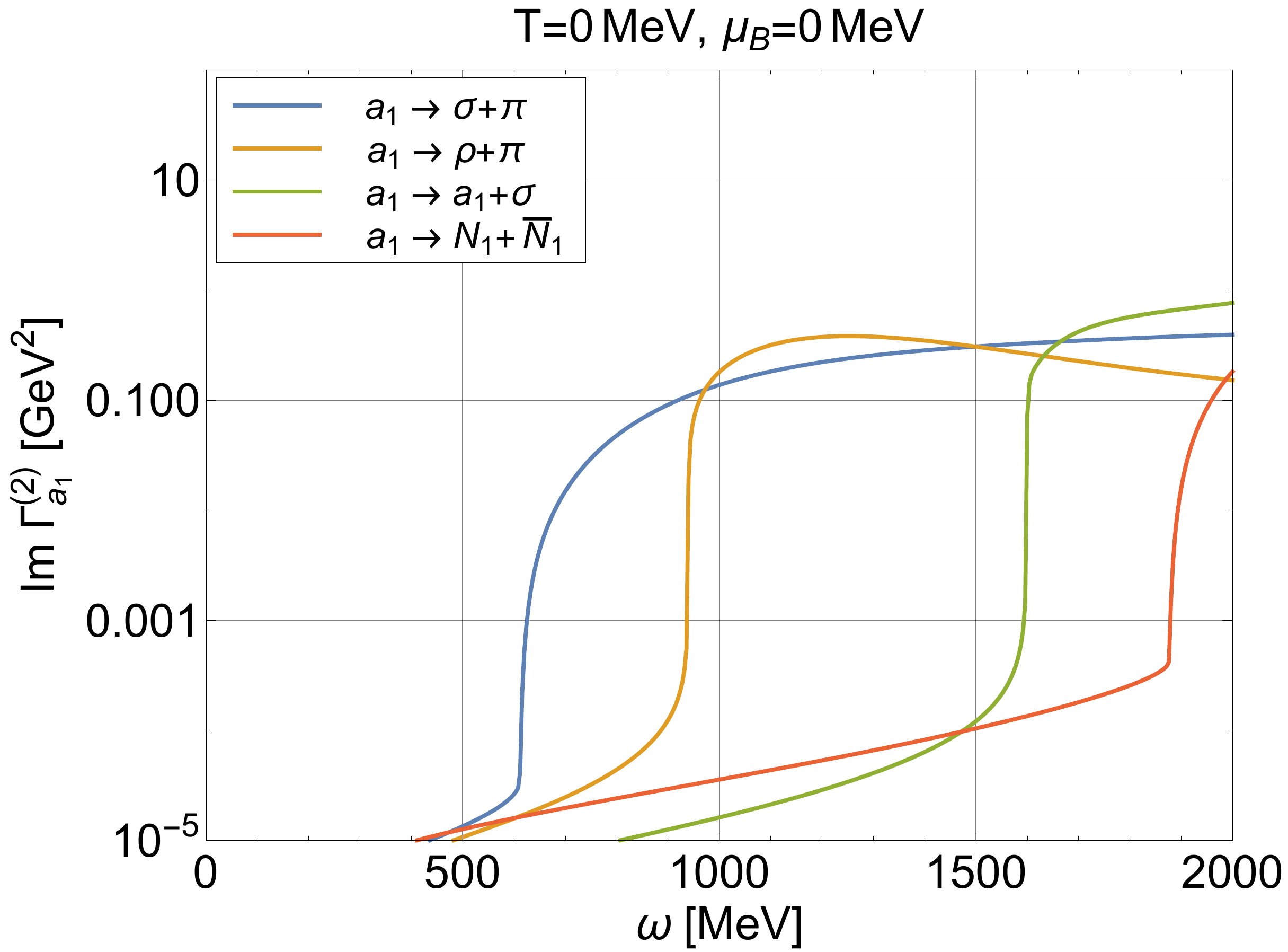}
	\includegraphics[width=\columnwidth]{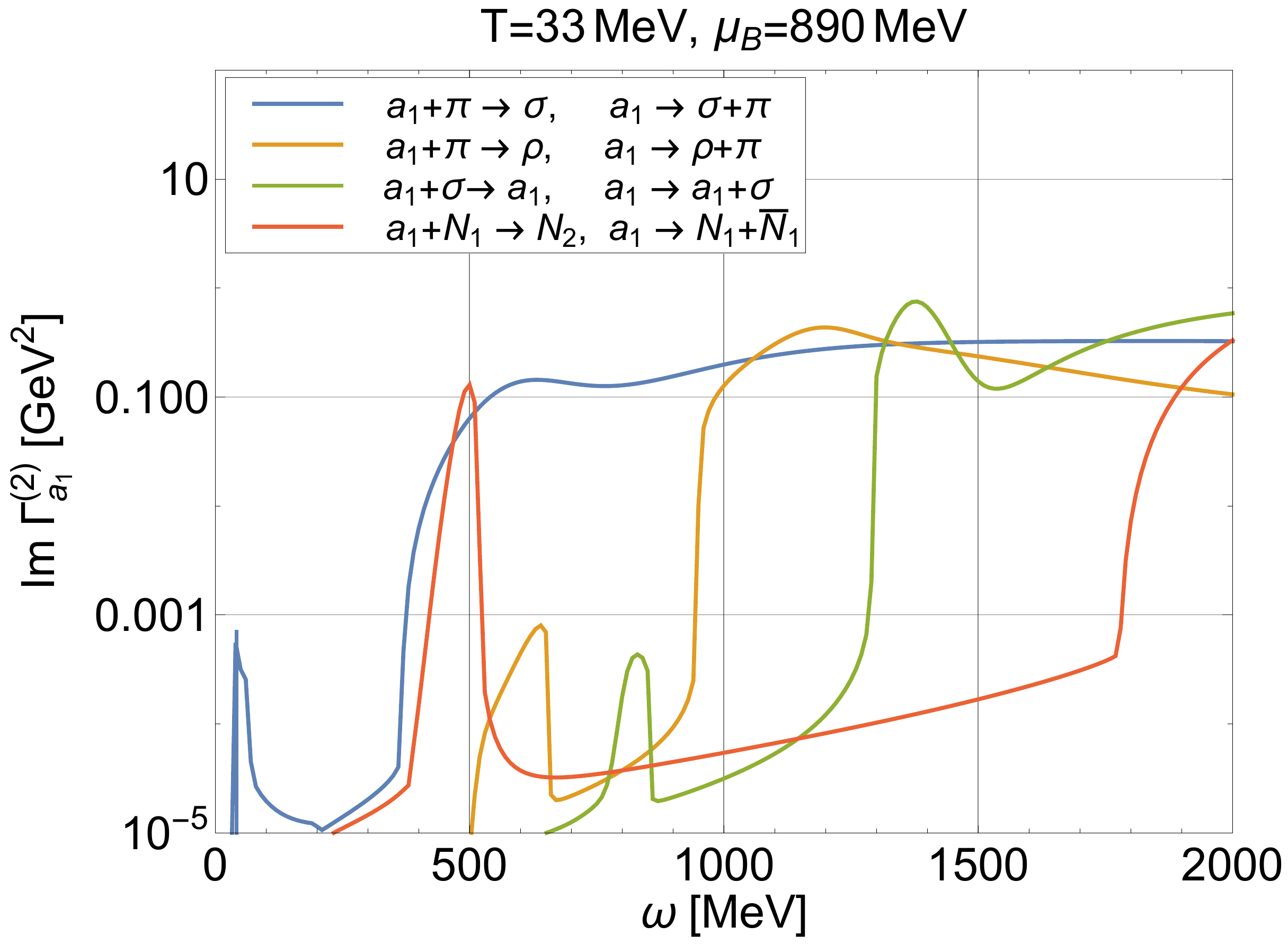}
	\caption{Imaginary part of the $\rho$ (top) and the $a_1$ (bottom) two-point function, evaluated at $\sigma_{0,\text{IR}}$ in the vacuum (left) and at $T=33$~MeV and $\mu_B=890$~MeV (right). The contributions from different loops are shown separately, cf.~Fig.~\ref{fig:2PF}. The decay thresholds are determined by the corresponding particle masses, cf.~Figs.~\ref{fig:masses_k} and \ref{fig:masses_mu}. See text for details.}
	\label{fig:ImGamma2}
\end{figure*}

So far, we have been working in Euclidean space-time. In order to calculate real-time quantities like retarded two-point functions and spectral functions we will now perform an analytical continuation of the flow equations for the two-point functions from imaginary to real energies using the previously introduced analytically-continued FRG (aFRG) technique \cite{Kamikado:2013sia,Tripolt:2013jra,Tripolt:2014wra}. This technique utilizes the fact that the FRG flow equations always have a one-loop structure and that therefore the well-known analytic continuation procedure for one-loop calculations can be applied, see e.g.~\cite{Das1997, Landsman:1986uw}.

To be more specific, the flow equations for the two-point functions, cf.~Eq.~(\ref{eq:2PF}) are analytically continued from imaginary to real energies by first using the periodicity of the bosonic and fermionic occupation numbers, which result from the Matsubara summation over the loop energy, w.r.t.~the discrete external Euclidean energy $p_0$, i.e. 
\begin{align}
n_{B,F}(E+\I p_0)\rightarrow n_{B,F}(E).
\end{align}
In a second step, the Euclidean energy $p_0$ is replaced by a continuous real frequency $\omega$ in the usual way,
\begin{align}
\Gamma^{(2),R}(\omega,\vec p)=-\lim_{\epsilon\to 0} \Gamma^{(2),E}(p_0=-\I(\omega+\I\epsilon), \vec p),
\end{align}
where we use a small but finite value of $\epsilon=0.1$~MeV in our numerical implementation. 

The resulting flow equations for the retarded two-point functions are then solved using the scale-dependent effective potential as well as the flow of the masses $m_{v,k}$ and $m_{0,k}$ evaluated at the IR minimum $\sigma_{0,\text{IR}}$ as input. The initial values of the retarded two-point functions are given by
\begin{align}
\Gamma^{(2),R}_{\rho,\Lambda}(\omega)&=m_{0,\Lambda}^2\left(1+\frac{m_{\rho,\Lambda}^2}{(\epsilon-i\omega)^2}\right),\\
\Gamma^{(2),R}_{a_1,\Lambda}(\omega)&=m_{0,\Lambda}^2\left(1+\frac{m_{a_1,\Lambda}^2}{(\epsilon-i\omega)^2}\right).
\end{align}
This initial shape will change as fluctuations are included by solving the flow equations. In particular, the two-point functions will show sudden changes and thresholds induced by the different processes possible in our setup. The ones representing a decay of an off-shell meson into two on-shell particles can occur even in the vacuum while the processes involving an additional particle in the initial state are only possible at finite temperature and/or density when a thermal medium of such particles is available. We note that the usual kinematic constraint for a decay process like $\rho^*\rightarrow \pi+\pi$ is given by $\omega \geq 2 m_\pi$ while for a thermal capture process like $a_1^*+\pi\rightarrow \sigma$ we must have $\omega \leq m_\sigma-m_\pi$.

For an off-shell rho-meson $\rho^*$ with energies $\omega$ up to 2~GeV the relevant processes here  are
\begin{align}
&\rho^*\rightarrow \pi+\pi, \\
&\rho^*\rightarrow a_1+\pi,
&&\rho^*+\pi\rightarrow a_1, 
\nonumber \\
&\rho^*\rightarrow N +\bar{N}, 
&&\rho^*+{N_1}\rightarrow N_2 .
\nonumber
\end{align}
Note that $\rho^*+a_1\rightarrow \pi $ is in principle also possible, at very large values of the baryon chemical potential, when eventually $m_\pi > m_{a_1}$, cf.~Fig.~\ref{fig:masses_mu}.
Other processes either require higher energies or involve anti-baryons in the heat bath which are exponentially suppressed. 
For the $a_1$-meson we have, analogously,
\begin{align}
&a_1^*\rightarrow \pi+\sigma,
&&a_1^*+\pi\rightarrow \sigma,
\\
&a_1^*\rightarrow \rho+\pi,
&&a_1^*+\pi\rightarrow \rho,
\nonumber\\
&a_1^*\rightarrow a_1+\sigma,
&&a_1^*+\sigma\rightarrow a_1,
\nonumber\\
&a_1^*\rightarrow N+\bar{N}, 
&&a_1^*+ {N_1}\rightarrow N_2. \nonumber
\end{align}
where again also the more exotic processes $a_1^*+\rho\rightarrow \pi$ and 
$a_1^*+a_1\rightarrow \sigma$ are possible in those regions of the phase diagram at very large $\mu_B$ where the (pseudo-)scalar masses increase beyond those of the (axial-)vectors.
More importantly, however, when the sigma mass drops below the pion mass, we can also have 
\begin{equation}
a_1^*+\sigma\rightarrow \pi \, .
\end{equation} 
This occurs only in the critical regions close to both CEPs, and it might hence serve as a potential signature of the existence of the chiral CEP, in particular.

In Fig.~\ref{fig:ReGamma2_vac} we show the IR result for the real parts of the $\rho$ and the $a_1$ two-point functions in the vacuum. We use the zero-crossings of these functions as an approximation for the pole masses of the respective resonances, which we fix to the phenomenological values of $m_\rho^p\approx 775$~MeV and $m_{a_1}^p\approx 1230$~MeV, cf.~\cite{ParticleDataGroup:2020ssz}, by adjusting the parameters to the values listed in Tab.~\ref{tab:pot_params} and in Sec.~\ref{sec:vector_masses}. We also observe the effects from different vacuum decay processes which give rise to sudden changes in the two-point functions. 

In Fig.~\ref{fig:ImGamma2} we show the imaginary parts of the $\rho$ and the $a_1$ two-point functions in the vacuum, as well as at $T=33$~MeV and $\mu_B=890$~MeV as a representative example for medium effects in a region where both endpoints might have an influence at the same time, that of the nuclear liquid-gas transition, here as a crossover from low to high density, as well as the chiral CEP approached at this temperature for higher $\mu_B$.

The imaginary part of the $\rho$-meson two-point function in the vacuum is practically zero below the lowest decay threshold which is determined by the process $\rho^*\rightarrow \pi+\pi$ and is located at $\omega\approx 280$~MeV. 

In particular for the nucleon-(anti)nucleon threshold, but also in $\rho^*\to a_1+\pi$, we observe the effects of the finite value for $\epsilon$ which produces small imaginary parts and hence results in non-zero values of the spectral function even below the thresholds. We note that the flow equations for the imaginary parts of the retarded two-point functions can in principle also be solved for $\epsilon=0$ exactly, see e.g.~\cite{Jung:2016yxl}. Here, however, we have used a finite value of $\epsilon$ in order to keep the numerical results consistent with the ones for the real parts, where the limit $\epsilon\rightarrow 0$ is not so straightforward to take numerically with the principal value prescriptions that are involved there. Moreover, a small value for $\epsilon$ has the additional benefit of mimicking a phenomenological two-particle-into-two-particle scattering continuum at arbitrarily low energies. At $T=33$~MeV and $\mu_B=890$~MeV, the imaginary part shows additional changes due to capture processes involving particles from the thermal medium, in particular from $\rho^*+\pi\to a_1$ and $\rho^*+ N_1 \to N_2$.

\begin{figure}[b!]
	\includegraphics[width=0.5\textwidth]{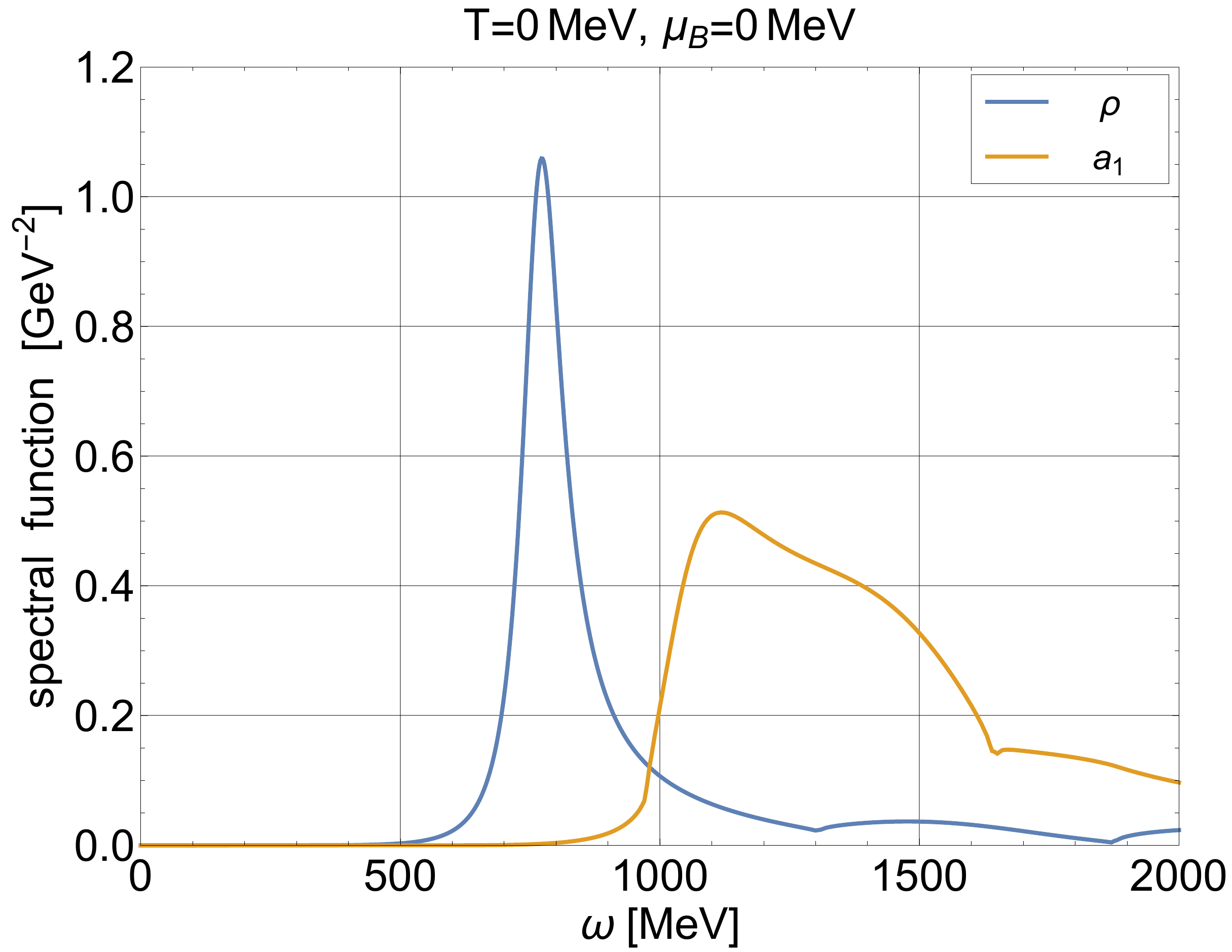}
	\caption{Spectral functions of the $\rho$ and the $a_1$ meson in the vacuum. The $\rho$ spectral function shows a prominent peak at its pole mass of $m_\rho^p\approx 775$~MeV while the $a_1$ spectral function exhibits a rather broad maximum which is strongly influenced by the decay $a_1\rightarrow \rho+\pi$, cf.~Fig.~\ref{fig:ImGamma2}.}
	\label{fig:spectral_vac}
\end{figure}

For the $a_1$ two-point function we observe in principle the same behavior as for the $\rho$ two-point function, especially in the vacuum where the lowest threshold starts at 
$a_1^*\rightarrow \pi +\sigma \approx 610$~MeV. This here represents the $\sigma$-resonance contribution to the three-particle decay $a_1 \to 3\pi $ which should of course start somewhat lower, at about 420~MeV, reflecting the smoother onset of the $3\pi$ continuum. To achieve this, one needs to feed the result for $\sigma $ two-point function as the broad $2\pi$ resonance back into the aFRG flow equations in a fully selfconsistent calculation in the future.

At finite $T$  and $\mu_B$ an increased number of processes and hence more complicated structures arise. In particular, we also observe a van Hove-like peak at $\omega\approx 35$~MeV in the contribution from $a_1^* +\pi\to \sigma $ which originates from an approximate saddle point in the difference of the corresponding scale-dependent quasi-particle energies  $E_{\sigma,k}-E_{\pi,k}$, see also \cite{Jung:2016yxl}. Another interesting effect is visible in the contribution from the $a_1^*\rightarrow a_1+\sigma$ process where a peak near the threshold is forming at $\omega\approx 1370$~MeV. This enhancement is due to the dropping sigma-meson mass in the crossover region above the nuclear liquid-gas transition. These structures, as all the effects visible in the imaginary parts of the two-point functions, directly translate into the shape of the spectral functions discussed in the following.

\subsection{Spectral functions}

\begin{figure*}[t!]
	\includegraphics[width=\columnwidth]{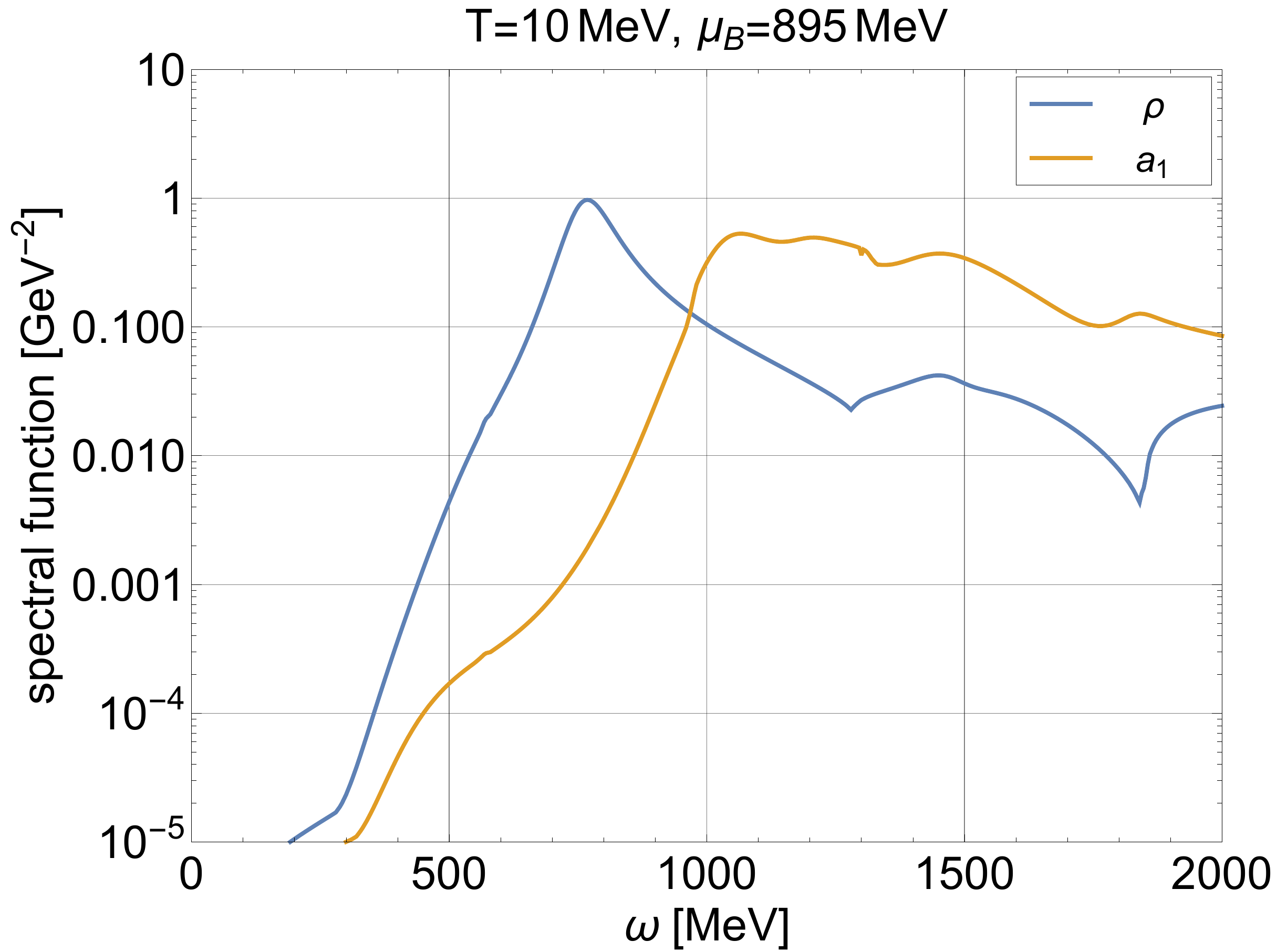}
	\includegraphics[width=\columnwidth]{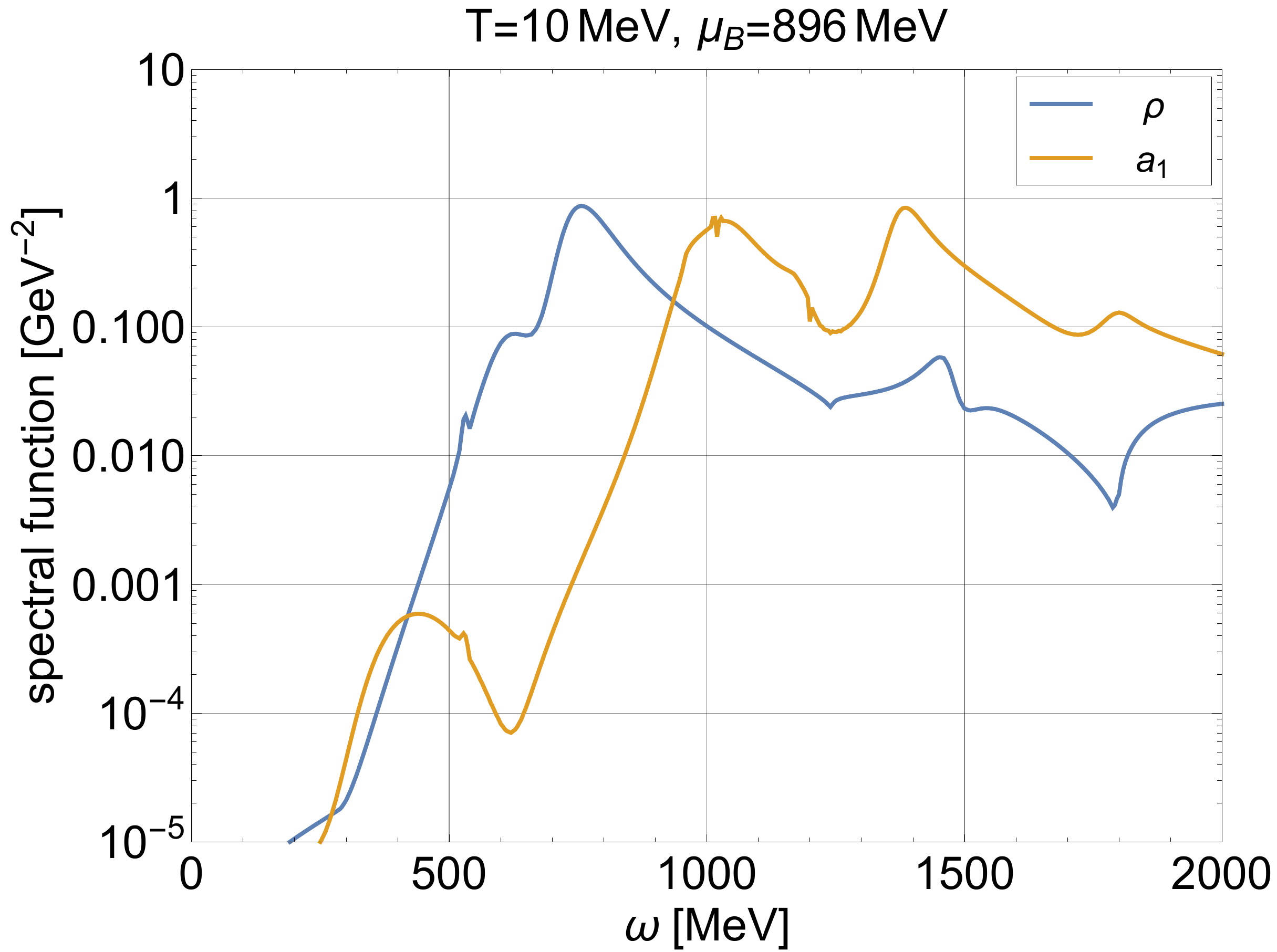}\\[2mm]
	\includegraphics[width=\columnwidth]{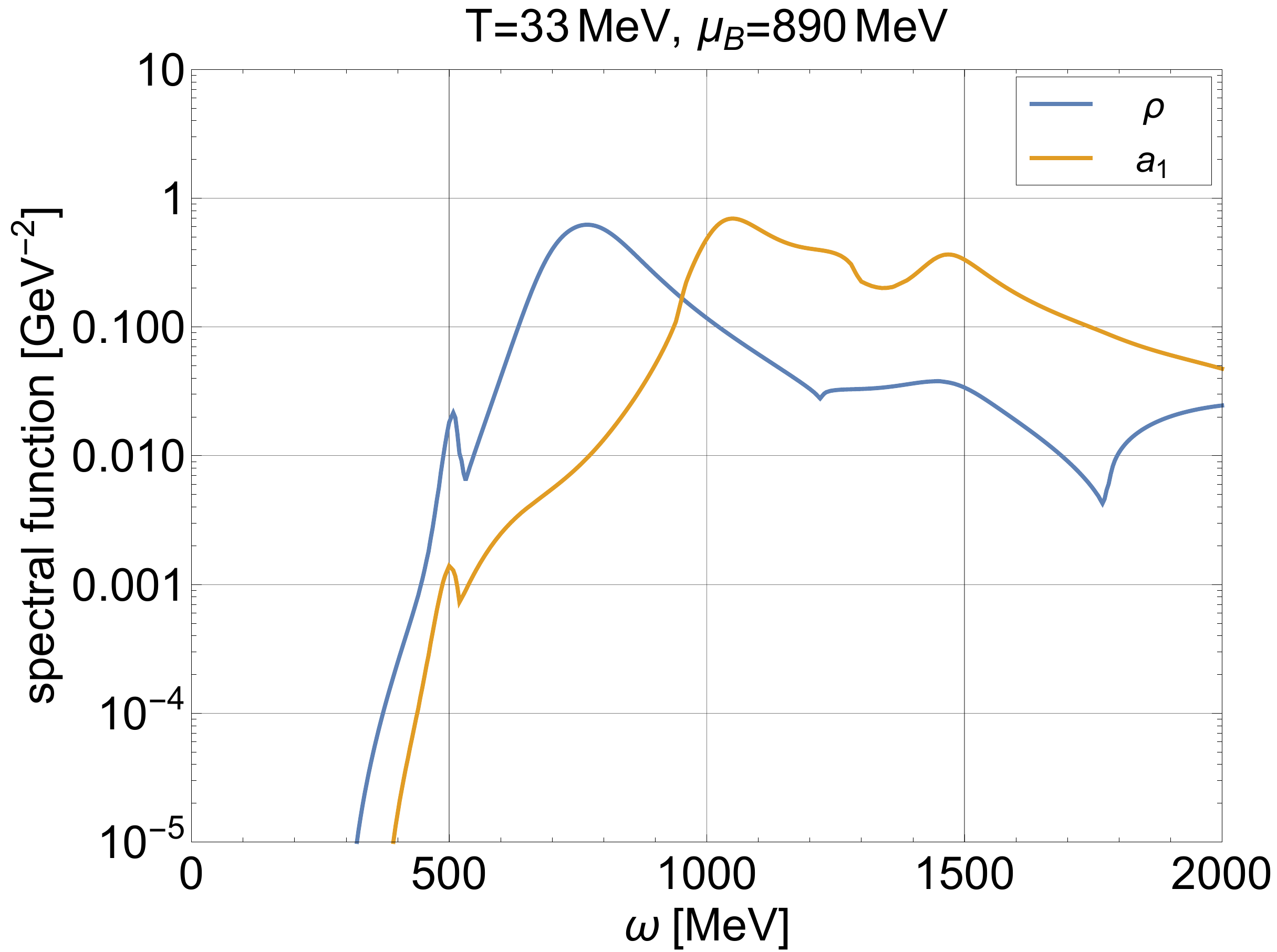}
	\includegraphics[width=\columnwidth]{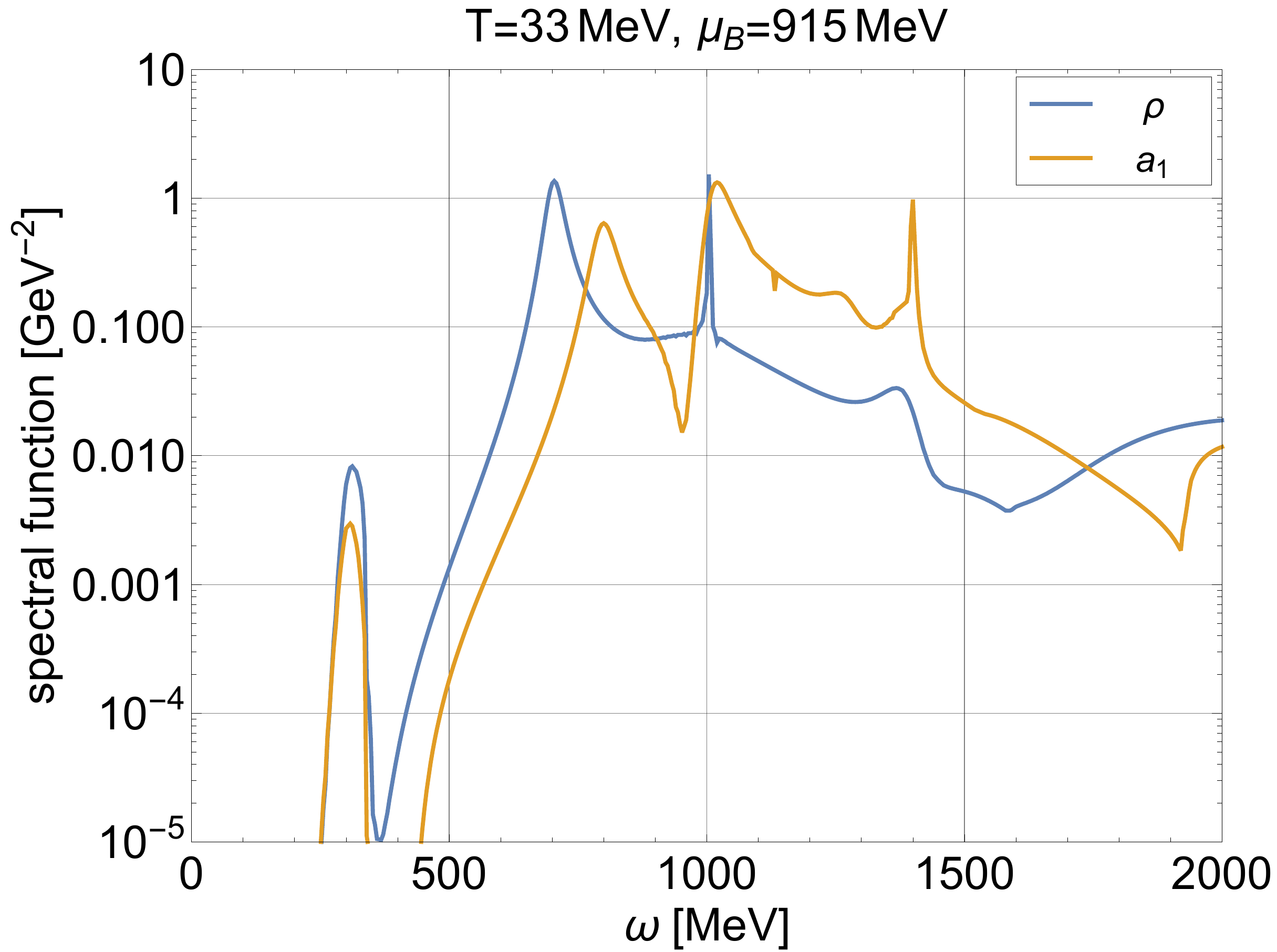}\\[-3mm]
	\caption{Spectral functions of the $\rho$ and the $a_1$ meson at different temperatures and chemical potentials. The spectral functions show complicated in-medium modifications due to the various decay and capture processes in the thermal medium. In particular near the CEP of the nuclear liquid-gas transition (top right) and in a regime approaching the chiral CEP (bottom right) we observe strong modifications. At $T=10$~MeV and $\mu_B=896$~MeV one sees the appearance of an additional peak structure in the $a_1$ spectral function at $\omega\approx 1380$~MeV which stems from the $a_1^*\rightarrow a_1+\sigma$ process, with the $\sigma$ meson encoding the critical behavior. At higher temperatures of $T=33$~MeV the effects from capture processes become more pronounced, e.g.~from the $a_1^*+N_1\rightarrow N_2$ process at $\omega\approx 500$~MeV (bottom left). At higher chemical potential (bottom right) we observe additional peak structures arising, this time also in the $\rho$ spectral function, as well as a progressing degeneration of the spectral functions due to the restoration of chiral symmetry. See text for details.}
	\label{fig:spectral_T}
\end{figure*}

\begin{figure*}[t!]
	\includegraphics[width=\columnwidth]{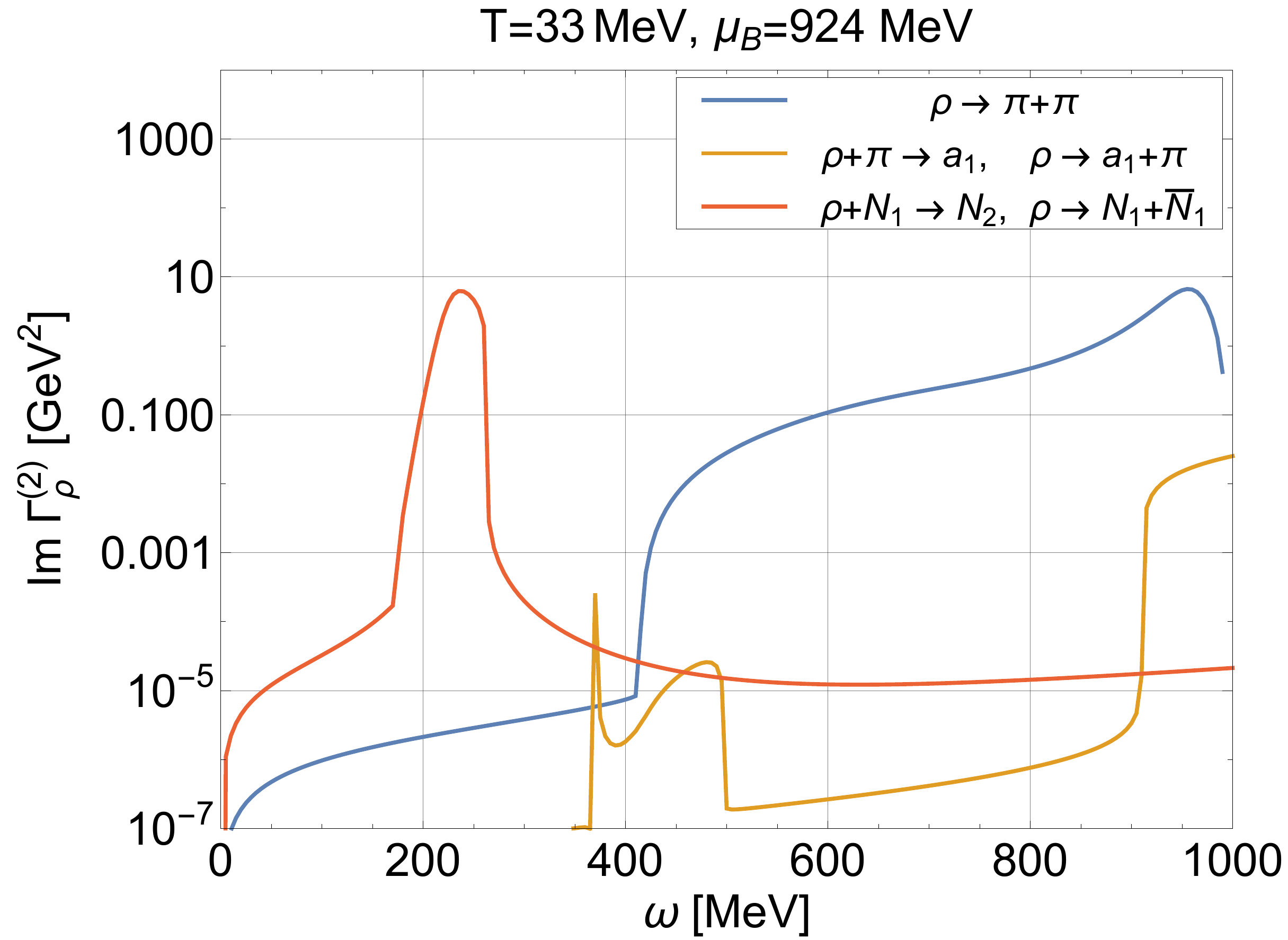}
	\includegraphics[width=\columnwidth]{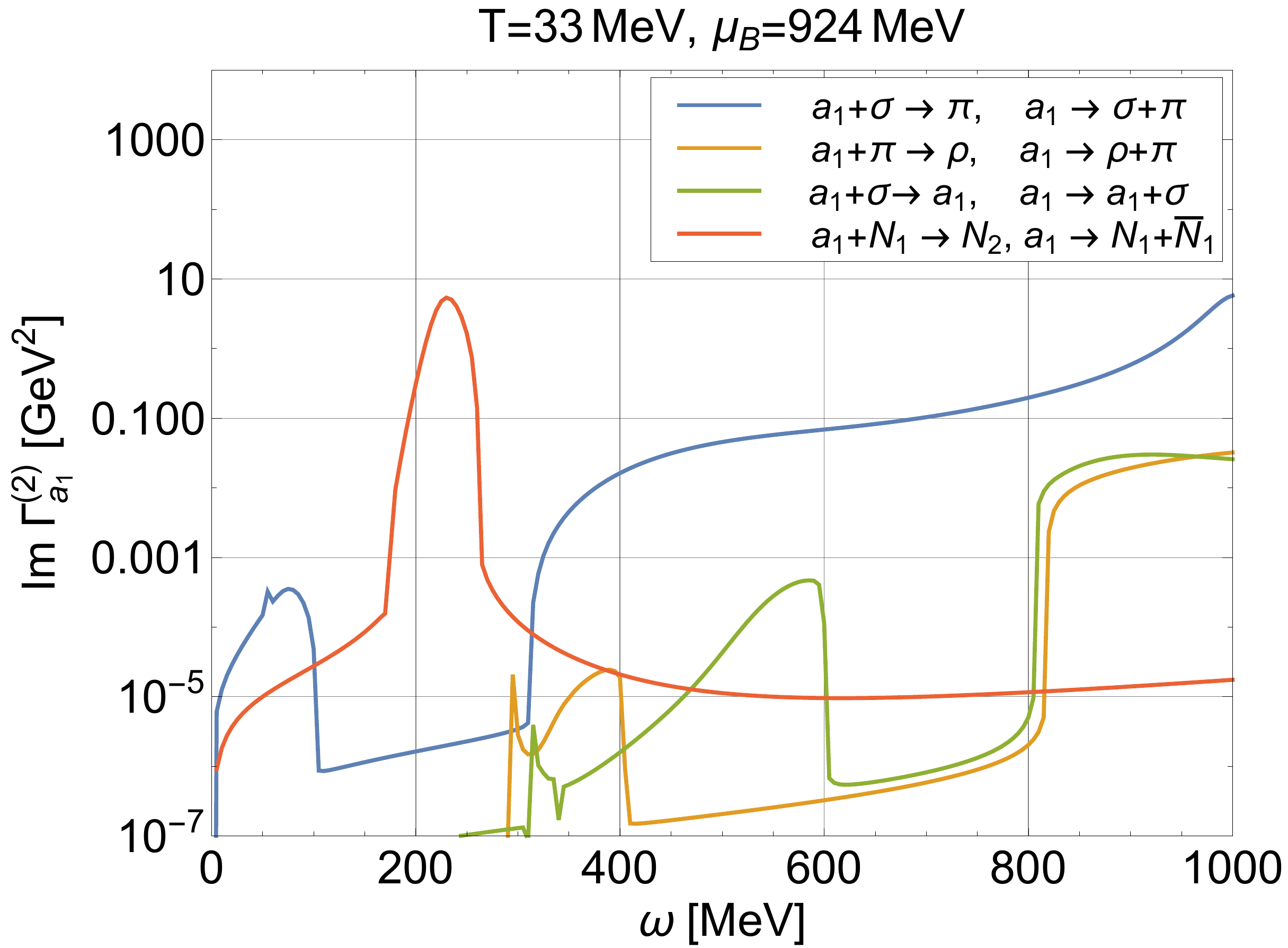}
	\caption{Imaginary part of the $\rho$ (left) and the $a_1$ (right) two-point functions at $T=33$~MeV and $\mu_B=924$~MeV, close to the chiral CEP. Here, a particularly small  value of $\epsilon=0.01$~MeV was needed in order to be able to resolve weak low-energy contributions from capture processes such as the critical $a_1^* +\sigma \to \pi $. As before, the separate components are extracted from the different loops shown in~Fig.~\ref{fig:2PF}.}
	\label{fig:ImGamma2_CEP}
\end{figure*}

\begin{figure}[t]
	\includegraphics[width=0.5\textwidth]{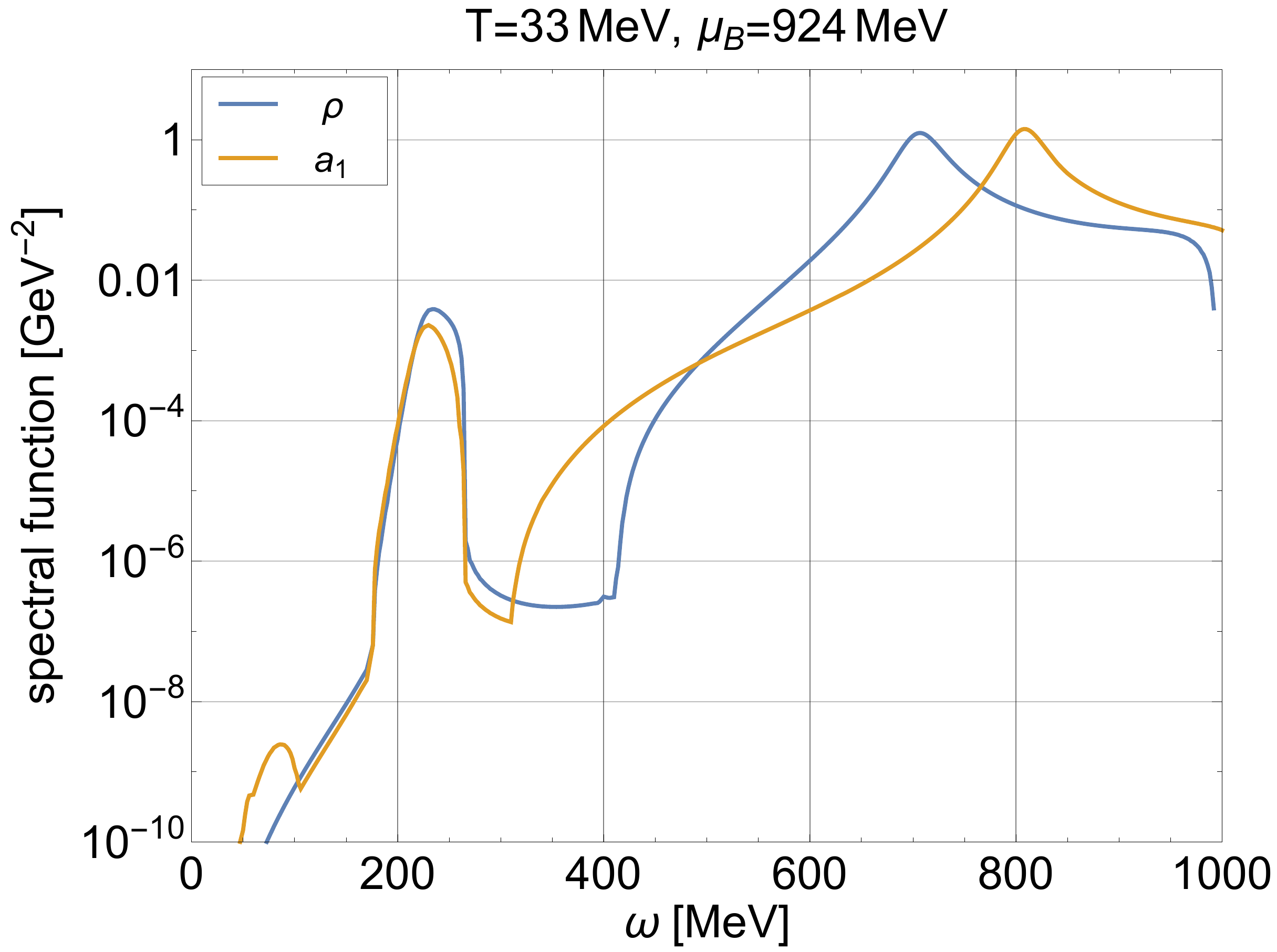}
	\caption{Critical spectral functions of the $\rho$ and the $a_1$ meson at $T=33$~MeV and $\mu_B=924$~MeV, close to the chiral CEP. The most prominent low-energy contributions to both spectral functions arise from baryon-resonance formation $\rho/a_1 + N_1 \to N_2$ which gives rise to prominent peaks around $\omega \approx 240$~MeV where the critical spectral functions have basically no support otherwise.}
	\label{fig:spectral_CEP}
\end{figure}

In this section we present our results for the in-medium $\rho$ and $a_1$ spectral functions in nuclear matter. The spectral function is generally defined as the discontinuity at the cut in the propagator along the timelike invariant-momentum axis and hence given by the imaginary part of the retarded Greens function $G^R$,
\begin{equation}
\rho(\omega,\vec p)=-\frac{1}{\pi}\text{Im}\, G^R(\omega,\vec p),
\end{equation}
which can be expressed in terms of the retarded two-point function as
\begin{equation}
\rho(\omega,\vec p)=\frac{1}{\pi}\frac{\text{Im}\,\Gamma^{(2),R}(\omega,\vec p)}{\left(\text{Re}\,
	\Gamma^{(2),R}(\omega,\vec p)\right)^2+\left(\text{Im}\,\Gamma^{(2),R}(\omega,\vec p)\right)^2}.
\end{equation}
In this work, we set the external spatial momentum $\vec{p}=0$ to zero which makes an additional splitting of the spectral functions into a part transverse and longitudinal to the medium unnecessary.

In Fig.~\ref{fig:spectral_vac} we show the $\rho$ and the $a_1$ spectral functions in the vacuum. The $\rho$ spectral function shows a prominent peak at its pole mass of $m_\rho^p\approx 775$~MeV and a full width of $\Gamma\approx 100$~MeV. 
The only process contributing in this energy regime is the decay into two pions, $\rho^*\rightarrow \pi+\pi$. Comparing to the experimental width  of 147.5~MeV for the charged $\rho $ into two pions \cite{ParticleDataGroup:2020ssz} our decay width is somewhat small but of the right order. 
At higher energies the decay channels $\rho^*\rightarrow a_1+\pi$ and $\rho^*\rightarrow N_1+ \bar N_1$ give rise to additional thresholds at around $1300$~MeV and $1880$~MeV, cf.~Fig.~\ref{fig:ImGamma2}. The $a_1$ spectral function shows a broad maximum between $\omega\approx 1000-1500$~MeV where the width is due to the processes $a_1^*\rightarrow \sigma+\pi$ and $a_1^*\rightarrow \rho+\pi$. At higher energies we observe the $a_1^*\rightarrow a_1+\sigma$ threshold while the $a_1^*\rightarrow N_1+\bar N_1$ contribution is very small below $\omega\approx 2$~GeV, cf.~Fig.~\ref{fig:ImGamma2}. We note in particular that this is the first time that the $\rho$ and $a_1$ spectral functions have been obtained within an aFRG setting without suffering from unphysical decay thresholds into quark-antiquark pairs. This is one of the reasons why we are using the hadronic effective theory which contains nucleons and their parity partners in the place of the quarks in chiral quark models such as the Nambu-Jona-Lasinio or quark-meson models, no matter whether these are enhanced by Polyakov-loop variables to model confinement or not.

In Fig.~\ref{fig:spectral_T} we show the $\rho$ and $a_1$ spectral functions at different temperatures and baryon chemical potentials. As also observed for the two-point functions, the spectral functions are essentially independent of $\mu_B$ at low temperatures until very close to the phase transition. The size of this `critical regime' is indeed very small, as evident from the large changes between $\mu_B=895$~MeV and the CEP at $\mu_B=896$~MeV at $T=10$~MeV. Here, we observe some effects from the capture processes 
$a_1^* +\pi \to \rho  $ and $a_1^*+N_1 \to N_2$
at lower energies as well as an additional peak structure forming in the $a_1$ spectral function at $\omega\approx 1380$~MeV. This effect is due to the process $a_1^*\rightarrow a_1+\sigma$ which is strongly affected by the dropping of the sigma mass at this second-order phase transition.

At $T=33$~MeV the effects from the capture processes can be observed more clearly, both at $\mu_B=890$~MeV and $\mu_B=915$~MeV, e.g.~near $\omega\approx 500$~MeV, mainly due to the baryonic capture processes $\rho^*/a_1^*+N_1\rightarrow N_2$. Closer to the chiral CEP, at $\mu_B=915$~MeV, we again observe the formation of an additional peak structure in the $a_1$ spectral function due to the critical effects entering via the process $a_1^*\rightarrow a_1+\sigma$. A similar peak structure is also visible in the $\rho$ spectral function which, however, originates from an additional zero-crossing forming in the real part of the two-point function which is connected to the process $\rho^*\rightarrow \pi+\pi$. We also observe that the $\rho$ and the $a_1$ spectral functions become increasingly degenerate due to the progressing restoration of chiral symmetry. In fact, one can show analytically that the flow equations of the $\rho$ and the $a_1$ two-point functions become degenerate in the limit $\sigma_0\rightarrow 0$. 

The most relevant low-energy processes at $T=33$~MeV and $\mu_B = 924$~MeV, near the chiral CEP, are shown in Fig.~\ref{fig:ImGamma2_CEP} where the different contributions to the imaginary parts of the $\rho$ and $a_1$ two-point functions are plotted up to 1~GeV (higher energies become increasingly difficult to compute as manifest in unphysical sign changes that can occur for higher energies in this critical region). Although we can clearly identify the potential signature of criticality in the process $a_1^* + \sigma\to \pi $ as mentioned above, the strength of this signal in the $a_1$ two-point function below $100$~MeV turns out to be very weak. The by far  dominant low energy features in both two-point functions here are the contributions from the nucleon capture processes $\rho^* + N_1 \to N_2 $ and  $a_1^* + N_1 \to N_2 $. These baryon-resonance formation processes give rise to rather strong peaks in the energy range around $\omega \approx 240$~MeV where there are no competing processes otherwise.

Finally, the critical low-energy behavior of the corresponding  $\rho$ and $a_1$ spectral functions at $\mu_B=924$~MeV and $T=33$~MeV, very close to the chiral CEP, is shown in Fig.~\ref{fig:spectral_CEP}.\footnote{The spurious sign changes in the imaginary parts of the two-point functions near criticality, as mentioned above, lead to positivity violations in the spectral functions at higher energies, above $1$~GeV. Whether these are related to thermodynamic instabilities observed in dense regions of the phase diagram with fluctuations  \protect\cite{Tripolt:2017zgc} remains to be investigated.} 

Of all contributions to the imaginary parts discussed above, the most prominent medium modifications of the critical spectral functions are the baryon-resonance formation processes $\rho^* + N_1 \to N_2 $ and  $a_1^* + N_1 \to N_2 $ which give rise to pronounced low-energy peaks around $\omega\approx 240$~MeV, below all other thresholds. The occurrence of these peaks is a unique prediction of the baryonic mirror assignment and its observation through enhanced dilepton pair production in the vicinity the chiral CEP would be an important confirmation of this picture of mass generation in QCD.
The critical capture process $a_1^* +\sigma \to \pi $ at even lower energies turns out by far too weak to be potentially significant. It is about six orders of magnitude lower than the baryonic capture process in the $a_1$ spectral function in Fig.~\ref{fig:spectral_CEP}. Other than that we observe only a small mass shift and broadening in the $\rho$ spectral function with considerably stronger medium modifications near the quasi-particle peak in the $a_1$, indicating the emerging restoration of chiral symmetry on the level of the eventually complete degeneration of the spectral functions of the chiral partners $\rho $ and $a_1$ at high density.

\section{Summary and Outlook}\label{sec:summary}

In the work presented here we discuss results on vector and axial-vector meson spectral functions at finite temperature and baryon-chemical potential, in order to assess the impact of chiral symmetry restoration in dense, low-temperature nuclear matter on the redistribution of spectral strength in both channels. As low-energy effective theory we use a chiral baryon-meson model, namely a parity-doublet model, which contains pions, sigma mesons, $\rho$ and $a_1$ mesons as well as nucleons and their parity partners chosen to be the $N^*(1535)$. Choosing hadronic degrees of freedom avoids unphysical quark-antiquark thresholds in the spectral functions in the confined phase. The vector and axial-vector mesons are introduced using a novel FRG formulation for massive vector fields based on (anti-)selfdual field strengths \cite{Jung:2019nnr}. In our opinion, this extended parity-doublet model captures the essential features of mass generation in QCD, in that hadron masses only partially result from the spontaneous breaking of chiral symmetry. On the other hand, the degeneracy in the spectral functions of parity partners in the restored phase is entirely driven by the evolution of the chiral condensate. The effects of thermal and quantum fluctuations are taken into account by using the FRG approach, which is known to describe phase transitions and critical phenomena in a way that is superior to a thermodynamic mean-field description mainly by including the dynamics of order-parameter fluctuations due to collective excitations.

Within this theoretical setup we have calculated the phase diagram for isospin-symmetric nuclear matter as a function of $T$ and $\mu_B$ as well as the screening masses of the various hadrons involved. The distinctive feature of the model is that it exhibits a nuclear liquid-gas phase transition as well as a chiral phase transition at a higher chemical potential where the nucleons and their parity partners become approximately degenerate but remain massive.
Similar to the chiral partners $\rho$ and $a_1$ in the vector-meson channel, the splitting between nucleons $N$ and their parity partners $N^*$ gradually disappears as chiral symmetry gets restored at finite density by predominantly the resonance mass dropping down to their common chirally invariant mass from the scale anomaly. Near the two first-order phase transitions as well as at their respective CEPs the Euclidean mass parameters of mesons and baryons all show the expected behavior and serve as input for the evaluation of the spectral functions.

For the calculation of the real-time two-point functions and the spectral functions we used the so-called aFRG method, of solving {\em analytically continued} FRG flow equations. Within this method one performs the analytic continuation from imaginary to real frequencies directly on the level of the FRG flow equations for the two-point functions and thus avoids the need for any numerical reconstruction. Moreover, it is thermodynamically consistent in that the effective potential and the spectral functions are calculated on the same footing. Using this approach, we have calculated the $\rho$ and the $a_1$ spectral functions at different temperatures and baryon-chemical potentials.

In the vacuum, the $\rho$ spectral function shows a prominent peak whose width is solely determined by the decay into two pions, as expected. The $a_1$ spectral function, in contrast, exhibits a very broad peak at higher energies which is determined by the decay into a pion and a sigma meson as well as into a rho meson and a pion, representing the $\sigma$ and $\rho$-meson resonance contributions to its three-pion decay width. For small temperatures and chemical potentials, below the liquid-gas phase transition, the spectral functions essentially coincide with those in the vacuum, as expected from the Silver-Blaze property. In the vicinity of the two critical endpoints, however, we observe significant changes with additional peaks emerging. Most strikingly are the modifications near the chiral CEP, were a prominent low-energy peak around 240~MeV shows up. Its origin can be traced back to the resonance excitation of the in-medium $N^*(1535)$ in both the vector and axial-vector channel. In fact, the excitation strength is nearly identical, reflecting the signature of parity doubling. Preliminary estimates indicate that this effect, which is strongest in the vicinity of the chiral CEP and possibly near the first-order boundary of the chiral phase transition might be observed experimentally in the vector channel through an increased dilepton yield at correspondingly low invariant masses measured in heavy-ion collisions at a few GeV/nucleon with high statistics. Its detection would yield strong evidence in support of the parity-doubling scenario as providing the mechanism for chiral symmetry restoration inside dense nuclear matter. 

To arrive at a satisfactory description of dense and warm nuclear matter and its spectral properties further improvements are called for. One relates to a quantitative description of the nuclear matter. For a phenomenologically acceptable description of the binding-energy per nucleon, the nuclear saturation density and the equation of state (EoS) of symmetric nuclear matter we will have to include the $\omega$ vector meson as a dynamical field in the calculation of the thermodynamic grand potential. From Walecka-type mean-field studies it is known that the repulsive nature of the $\omega$ meson is essential for a realistic nuclear matter EoS. Possible further improvements include taking into account higher truncation orders in the effective average action or using a self-consistent setup where the spectral functions are back-coupled into flows of effective potential and two-point functions. 
In addition, a calculation of the expected dilepton yields in heavy-ion collisions at a few GeV/nucleon is left for future work. Other phenomenologically important extensions will include the EoS of highly isospin-asymmetric nuclear matter and determination of neutrino emissivities relevant for binary neutron-star merger events from the corresponding weak (axial-)vector spectral functions in warm and dense neutron matter.

\acknowledgments
R.-A.~T. is supported by the Austrian Science Fund (FWF) through Lise Meitner grant M 2908-N. This work was supported by the Deutsche Forschungsgemeinschaft (DFG) through the grant CRC-TR 211 ``Strong-interaction matter under extreme conditions'' and the German Federal Ministry of Education and Research (BMBF) through grants No.~05P18RFFCA and No.~05P18RGFCA. 

\appendix

\section{Explicit Expressions}
\label{app:explicit}

In this appendix we summarize explicit expressions for various quantities used in this work. We begin with the regulator functions, for which we use three-dimensional Litim-type regulators \cite{Litim:2001up}, which allow for an analytic evaluation of Matsubara sums. More explicitly, we employ the following regulator functions for (pseudo-)scalar mesons, (axial-)vector mesons, and nucleons,
\begin{align}
R_{\sigma/\pi,k}(p)&=(k^2-\vec{p}^2)\Theta(k^2-\vec{p}^2),\\
R_{\rho/a_1,k}^{T,L}(p)&=-\frac{m_{0,k}^2}{p^2}(k^2-\vec{p}^2)\Pi_{\mu\nu}^{T,L}(p)\Theta(k^2-\vec{p}^2),\\
R_{N}(p)&=-i \slashed{\vec{p}}\left(\sqrt{k^2/\vec{p}^2-1}\right)\Theta(k^2-\vec{p}^2).
\end{align}
The corresponding regulator shape function is given by
\begin{align}
r(y)=\left( \frac{1}{y}-1\right)\theta(1-y),
\end{align}
with $y=p^2/k^2$.

Upon evaluating the (unregulated) Matsubara sums over internal energies in the various flow equations, we encounter the bosonic and fermionic occupation number factors which are given by
\begin{align}
n_B(E)&=\frac{1}{e^{E/T}-1},\\
n_F(E)&=\frac{1}{e^{E/T}+1}.
\end{align}
For the four-dimensional adjoint representation of the two-flavor $SU(2)_L \times SU(2)_R $ chiral symmetry we use $O(4)$-vectors $\phi = (\sigma , \vec \pi)^T$ and hermitian generators $\vec T_{L,R}$,
\begin{align}
    iT_L^a = \frac{1}{2} 
    \begin{pmatrix}
     0 & -\vec e_a^{\,T}\\
     \vec e_a & \varepsilon_{aij} 
    \end{pmatrix} , \;
    iT_R^a = \frac{1}{2} 
    \begin{pmatrix}
     0 & \vec e_a^{\,T}\\
     -\vec e_a & \varepsilon_{aij} 
    \end{pmatrix} ,
\end{align}
with $a,i,j \in \{1,2,3\}$ and $(\vec e_a)_i = \delta_{ai} $. The corresponding (axial-)vector martices $\vec T_{V,A} = \vec T_R \pm \vec T_L $ are then given by
\begin{align}
    iT_V^a = 
    \begin{pmatrix}
     0 &  0 \\
     0 & \varepsilon_{aij} 
    \end{pmatrix} , \;
    iT_A^a = 
    \begin{pmatrix}
     0 & \vec e_a^{\,T}\\
     -\vec e_a & 0 
    \end{pmatrix}\, .
\end{align}
The minimal coupling between (pseudo-)scalar and (axial-)vector mesons is then defined by the covariant derivative $D_\mu = \partial_\mu + i g V_\mu $ with the matrix valued vector field
$V_\mu = \vec\rho_\mu \vec T_V + \vec a_{1\mu} \vec T_A$. Explicitly, this leads to 
\begin{align}
    \frac{1}{2}  (D_\mu \phi)^\dagger D_\mu \phi &= \\
    & \hskip -1cm \frac{1}{2}  (\partial_\mu \phi)^T \partial_\mu \phi + ig   (\partial_\mu \phi)^T V_\mu \phi + \frac{g^2}{2} \phi^T (V_\mu)^2 \phi \, .\nonumber
    \\[-10pt] \nonumber
\end{align}
The three-point interactions of order $g$ then yield,
\begin{align}
 i (\partial_\mu \phi)^T V_\mu \phi &=\\[4pt]
 &\hskip -1cm (\vec\pi \times \vec{\rho}_\mu) \!\cdot\! \partial_{\mu}\vec{\pi}  -\sigma\vec{a}_{1\mu}\!\cdot\! \partial_{\mu}\vec{\pi}+\vec{a}_{1\mu}\!\cdot\!  \vec{\pi} \partial_{\mu}\sigma \, ,\nonumber %
 \end{align}
and the quartic interactions of order $g^2$ become 
\begin{align}
\frac{1}{2} \phi^T (V_\mu)^2 \phi &=\\
&\hskip -.1cm
\frac{1}{2} (\vec{\rho}_\mu\times\vec{\pi}-\sigma\vec{a}_{1\mu})^2+ \frac{1}{2} (\vec{a}_{1\mu}\!\cdot\!\vec{\pi})^2\, .\nonumber
\end{align}
Finally, we list the explicit expressions for the three-and four-point vertices which can be obtained by taking three and four functional derivatives of the ansatz for the effective average action with respect to the various fields, cf.~Eq.~(\ref{eq:Gamma}). From the 
relevant contributions to the effective average action listed explicitly above,
 we obtain the three-point vertex functions involving vector mesons used in this work as
\begin{align}
\Gamma^{(3)}_{\rho^i_{\mu}\pi^j\pi^k}(q^{j},q^{k})&=
ig\varepsilon_{ijk}\,(q^k_{\mu}-q^j_{\mu})\, ,\\
\Gamma^{(3)}_{\sigma a_{1\mu}^i\pi^j}(q,q^j)&=
ig\delta_{ij}\, (q_\mu-q^{j}_\mu)\, ,\\
\Gamma^{(3)}_{\sigma a_{1\mu}^i  a_{1\nu}^j}&=
2g^2\sigma_0^2 \, \delta_{\mu\nu}\, \delta_{ij}\, ,\\
\Gamma^{(3)}_{ a_{1\mu}^i\rho^{j}_\nu \pi^k}&=
-g^2\sigma_0^2 \, \delta_{\mu\nu}\, \varepsilon_{ijk}\, ,
\end{align}
where all momentum arguments denote the incoming momenta of the (pseudo-)scalar mesons, $q_\mu$ for the sigma meson and $q_\mu^i$ for the isovector pions $\pi^i$. 
The four-point vertices are analogously obtained as
\begin{align}
\Gamma^{(4)}_{\rho^{i}_\mu \rho^{j}_\nu\pi^k\pi^l } &=
g^2 \delta_{\mu\nu}\, (2\delta_{ij}\delta_{kl}-\delta_{ik}\delta_{jl}-\delta_{il}\delta_{jk})\, ,\\
\Gamma^{(4)}_{ a_{1\mu}^i a_{1\nu}^j  \pi^k\pi^l }&=
g^2 \delta_{\mu\nu}\, (\delta_{ik}\delta_{jl}+\delta_{il}\delta_{jk})\, ,\\
\Gamma^{(4)}_{\sigma\sigma a_{1\mu}^i a_{1\nu}^j  }&=
2 g^2\delta_{\mu\nu} \,  \delta_{ij}\, .
\end{align}
The three-point couplings of the (axial-)vectors to the two baryons doublets $N_d$, $d= 1,\, 2$, are readily obtained from the ansatz for the effective average action and are given by
\begin{align}
\Gamma^{(3)}_{\rho_\mu^i \bar{N}_{d}N_{d}} &=ih_{v,d}\,\gamma_\mu \tau^i\, ,\\
\Gamma^{(3)}_{a_{1\mu}^i \bar{N}_{d}N_{d}} &=ih_{v,d}\, \gamma_\mu\gamma^5 \tau^i\, .
\end{align}

\end{document}